# Biology of Prions

Verónica Inés Cacace
Instituto de investigaciones cardiologicas, Universidad de Buenos Aires-Conicet
Buenos Aires
Argentina

Jun, 2011

Corresponding autor: veroinesc@yahoo.com.ar

## Abstract

In this paper we will review various aspects of the biology of prions and focused on what is currently known about the mammalian PrP prion. Also briefly describe the prions of yeast and other fungi.
Prions are infectious proteins behaving like genes, ie proteins that not only contain genetic information in its tertiary structure, ie its shape, but are also able to transmit and replicate in a manner analogous to genes but through very different mechanisms. The term prion is derived from "proteinaceous infectious particle" and arose from the Prusiner hypothesis that the infectious agent of certain neurodegenerative diseases was only in a protein, without the participation of nucleic acids. Currently there are several known types of prion, in addition to the originally described, which are pathogens of mammals and they has found in yeast and other fungi too. Prion proteins are ubiquitous and not always detrimental to their hosts. This vision of the prion as a causative agent of disease is changing, finding more and more evidence that could have important roles in cells and contribute to the phenotypic plasticity of organisms through the mechanisms of evolution.

## Resumen

En presente trabajo se hara una revision sobre diversos aspectos de la biología de priones, haciendo incapie en lo que actualmente se conoce acerca del prion PrP de mamiferos. Ademas se describiran brevemente los priones de levaduras y otros hongos.
Los priones son proteínas infecciosas que se comportan como genes, esto es, proteínas que no solo contienen información genetica en su estructura terciaria, es decir en su conformación, sino que también son capaces de transmitirla y replicarla en forma análoga a los genes pero por medio de mecanismos muy distintos. El termino prion proviene de "proteinaceous infectious particle" y surgio a partir de la hipótesis de Prusiner de que el agente infeccioso de ciertas enfermedades enfermedades neurodegenerativas consistía unicamente en una proteína, sin participación de acidos nucleicos. Actualmente se conocen varios tipos de prion, ademas de los descriptos originalmente, que son patógenos de mamíferos y tambien se los ha encontrado en levaduras y en otros hongos. Se trata de proteínas ubicuas y no siempre perjudiciales para sus hospedadores. Esta vision del prion como un agente causante de enfermedad esta cambiando, al encontrarse cada vez mas evidencia de que pueden cumplir funciones importantes en las células y contribuir a la plasticidad fenotipica de los organismos a traves de los mecanismos de la evolución.

## Introducción:

Los priones son proteínas infecciosas que existen en más de un estado conformacional, es decir, pueden adoptar distintas conformaciones en distintas condiciones, por lo cual, son excepciones a la teoría clásica del plegamiento de las proteinas, que afirma que la secuencia de aminoácidos determina un único plegamiento de la proteína, es decir,su conformación o estructura terciaria .
Ademas, los priones llevan información en dicha conformación, de manera análoga a los genes. Cada conformación tiene una información distinta que es heredable. Se comportan no solo como agentes infecciosos sino también como genes. Pueden transmitir, replicar y expresar en forma de algun fenotipo la información que contienen.

Se conocen varios tipos de prion, en mamíferos se encuentra el prion prp, responsable de las encefalopatias espongiformes transmisibles (TSEs), que son enfermedades neurológicas invariablemente letales y que actualmente no tienen tratamiento. Este prion puede existir en dos conformaciones: prpsc y prpc. Prpsc es la version infecciosa, que causa TSE, mientras que prpc es la versión normal de la proteina y no produce enfermedad, al contrario cumpliria funciones celulares normales. Este prion esta altamente conservado en la evolución y estaría presente no solo en mamíferos sino también en la mayoría de los grupos de vertebrados, por ejemplo se lo encontro en el pollo y en reptiles (1,2). Hasta donde se conoce, solo en mamíferos produce enfermedad.

Los demas priones conocidos actualmente son los de levaduras y hongos filamentosos.
Además, recientemente se detectó un nuevo prion en un molusco marino (genero Aplyssia) que cumple funciones relacionadas con la formación y mantenimiento de la memoria, lo cual es muy novedoso (3).

En Saccharomices cerevisiae se encuentran los priones que forman amiloides, denominados URE3, PSI y PIN (4,5). Además, actualmente, la proteasa vacuolar beta de levadura también es considerada un prion. La versión normal de esta proteína es simplemente una proteasa que degrada proteínas innecesarias. La versión infecciosa es denominada simplemente beta y activa a su propio precursor, de esta forma se autorreplica comportándose como prion (6). Por otra parte, en el hongo filamentoso Podospora anserina se encuentra el prion Het-s (7).

Los priones Het-s y beta cumplen funciones celulares: Het-s interviene en la incompatibilidad citoplasmática mientras que beta como fue indicado, participa de la degradación de proteínas innecesarias. URE3 y PSI son infecciosos, se comportan en forma análoga al prp de mamíferos. Aparentemente, PIN es neutral (no cumple funciones celulares ni produce enfermedad) (4,5,6,7).

Todos los priones son proteínas codificadas en el genoma del organismo hospedador. Invariablemente presentan más de una conformación y cada una de ellas otorga una característica fenotípica distinta. Tienen la capacidad de autopropagarse en forma infecciosa y de ser heredables. Además, en todos los casos (excepto en el prion beta), la conformación infecciosa es capaz de formar agregados del tipo amiloide proteasa resistentes (6).

En cuanto a la nomenclatura, se denomina prion a la conformación infecciosa de una dada proteina, la que tiene características autorreplicativas y se nombra con otro nombre a la conformación considerada normal de la misma. Por ejemplo, en el caso de la proteina prp, se toma como prion a la conformación prpsc y a prpc se lo considera simplemente una proteína normal no prionica. De esta manera, es comun encontrar en la bibliografía frases como por ejemplo que prpsc es un prion de prpc o que Psi es un prion de la proteina Sup35.
No hay que olvidar que aunque se utilice esta nomenclatura, en realidad las conformaciones son versiones distintas de una misma proteína (con una única secuencia de aminoácidos). Por esto, en principio se podrían considerar como prionicas a todas ellas y la diferencia estaria dada por el hecho de que algunas conformaciones son autorreplicativas comportándose como priones clásicos y otras cumplen otra funcion. De esta forma, por ejemplo prp es un prion mientras que prpsc y prpc son las posibles conformaciones que puede tener.
En sentido amplio, tal vez se deberia definir como prion a cualquier proteina con distintas conformaciones cada una con diferente funcion y capaz de cambiar de una conformación a otra.

En presente trabajo se hara una revision sobre diversos aspectos acerca de la biología de priones, haciendo incapie en lo que actualmente se conoce acerca del prion PrP de mamiferos. Ademas se describiran brevemente los priones de levaduras y otros hongos.

## El prion prp

La proteina prp fue identificada por primera vez en experimentos que intentaban demostrar la presencia de ácidos nucleicos virales como componentes del agente infeccioso de las encefalopatias espongiformes transmisibles (TSEs). El agente causante de las mismas, purificado parcialmente del cerebro de animales enfermos de TSE resultó ser una proteína insoluble de 33 kD que fue denominada prpsc que significa prion del scrapie por el nombre de la TSE estudiada que era el scrapie, que afecta al ganado ovino. Pronto se observo que el tratamiento de esta proteína con proteinasa K generaba un fragmento de 27-30 Kd que fue identificado como el principal componente del agente infeccioso. Este fragmento resulta ser infectivo cuando es inoculado en animales de laboratorio (8).
Con técnicas de secuenciacion se obtuvo la secuencia aminoacídica de la porción N-terminal de la proteina prpsc y con esta fue posible sintetizar la secuencia de nucleótidos correspondiente que se utilizó como sonda para encontrar el supuesto gen de origen viral.  Este gen se encontro donde menos se lo esperaba: en el genoma del hospedador y no correspondia a un gen viral sino a un gen cromosomal copia unica ubicuo en las distintas especies de vertebrados (1,2) independientemente de la existencia o no de infección por prpsc (9).

Dicho gen fue denominado PRP (tambien conocido como PRNP) y su producto genico es la proteina prp, mayormente la versión normal (denominada prpc, por “cellular prion protein”. Esta proteina puede originar una isoforma patogénica (denominada prpsc) de forma espontanea si su gen PRP presenta alguna mutación puntual que favorezca un cambio de conformación.  La presencia de prpc es totalmente necesaria para que la infección por priones ocurra. Los ratones knockout para el gen PRP son resistentes a la infección con priones prpsc inoculados por  diversas vias como la periferica o la intracerebral (9).

La secuencia de aminoácidos de prpc y de prpsc por supuesto es la misma, como ya fue indicado anteriormente, ambas son la misma proteina pero con distinta conformación. Se las puede llamar confórmeros o isoformas.  Las diferencias entre ambas isoformas radican en el hecho de que prpc presenta un 40% de $\alpha$-helices y menos del 10% de láminas $\beta$ en su estructura terciaria mientras que prpsc tiene alrededor de un 50% de láminas $\beta$ (figura 1).  La capacidad de prpsc de ser infecciosa radica en su conformación (8,10), se trata de una versión alterada de prp que perdio su funcion fisiológica normal pero gano la capacidad de autorreplicarse  mediante la conversión de moléculas de prpc (que utiliza como molde) en mas moléculas de si misma. Otra característica que permite distinguir a prpc de prpsc es la resistencia al tratamiento con proteasas: prpc es totalmente degradada pero prpsc resulta ser parcialmente resistente a estas enzimas.  El tratamiento de prpsc con proteinasa K produce un peptido de 142 aa denominado prp 27-30 que tiene la capacidad de polimerizar en amiloides y es neurotóxica (10).

Prpc está altamente conservada en la evolución. La similitud entre la Prpc de distintas especies de mamíferos oscila de un 85% al 97%, mientras que al comparar la secuencia entre primates y humanos la similitud es del 92,9 -99,6 %  (11).

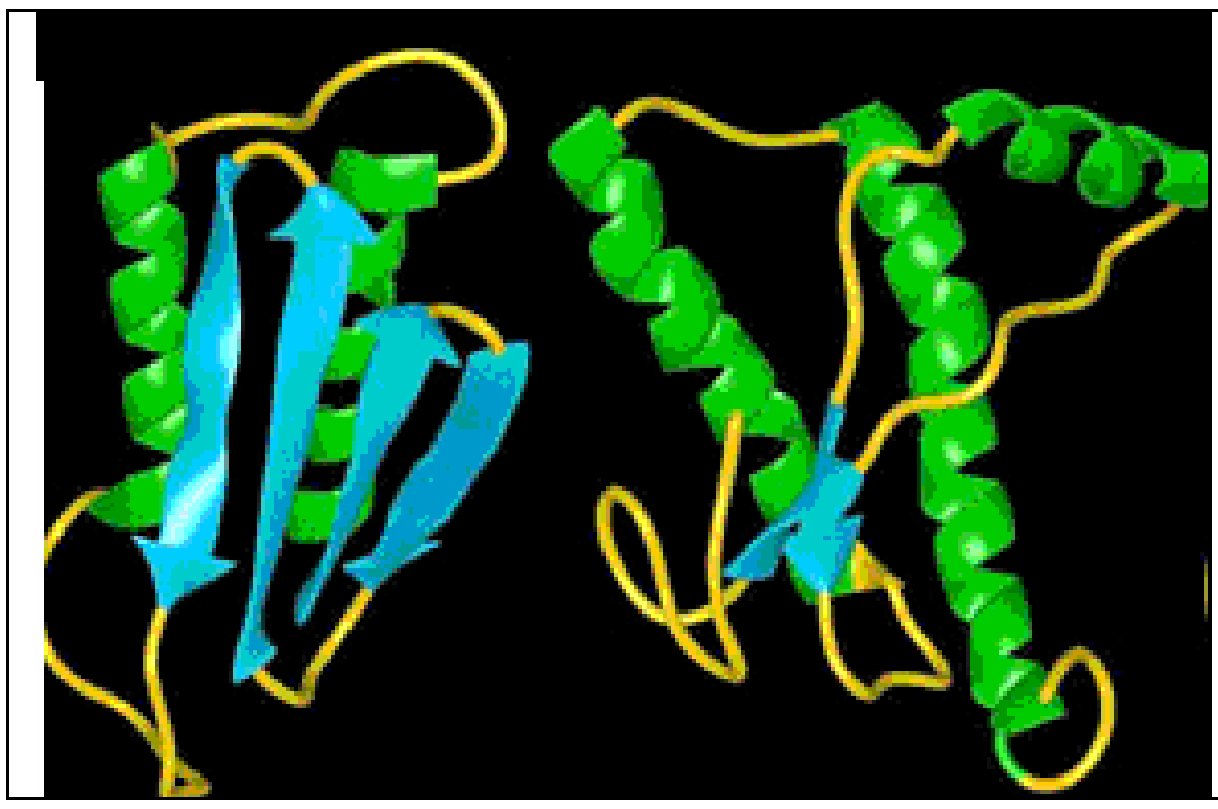

Figura1.Estructurasproteicas obtenidas con Rasmol. A la derecha se muestra la estructura de prpc, mientras que la imagen de la izquierda corresponde a prpsc. Se observa el mayor contenido de láminas $\beta$ en prpsc, mientras que en prpc predominan las $\alpha$-helices.

Prpc es una proteína celular normal que se expresa en las neuronas y células de la glía en el cerebro y médula espinal. También está presente en otros tejidos, principalmente tejido linfoide como el bazo y en las células del sistema inmune tales como células dendríticas y linfocitos B (12,13,14,15).
El mensajero de prpc aparece en el cerebro de raton y de pollo en etapas tempranas del desarrollo embrionario y su nivel aumenta en el transcurso de la embriogénesis (13,14). En el cerebro adulto, el mRNA de prp esta amplimente distribuído y se concentra en zonas tales como las células de Purkinje del cerebelo, neuronas hipocampales, neocorticales y motoras (16).

## Estructura de prpc:

El gen humano PRP está situado en el brazo corto del cromosoma 20. Posee un solo exón a diferencia de los correspondientes a ratones, vacas y ovejas que tienen 2 exones. Los promotores de los genes de hamster y de ratón poseen varias copias de secuencias repetidas ricas en G+C pero carecen de TATA box (17,18).

El gen PRP codifica a prpc, que posee alrededor de 250 aminoacidos según la especie. El transcripto primario de prpc no presenta splicing alternativo (19).

La proteína prpc consta de dos dominios: una porcion N-terminal flexible capaz de unir iones cobre, con una serie de 5 repeticiones de octapeptidos ricos en prolina y glicina y una porcion C-terminal con un segmento hidrofóbico central altamente conservado que posee tres $\alpha$-helices (denominadas H1, H2, H3 ) (20). Además en el extremo C-terminal posee una secuencia para la adición de un ancla de fosfatidilinositol (GPI) mientras que en el extremo N-terminal posee un péptido señal de 22 aa (KDEL) para dirigir la biosíntesis de la proteína al retículo endoplásmico. Dicho péptido es clivado una vez que prpc es translocada al retículo (21).

Por otra parte, en lo que respecta a la conformación prpsc, como se indicó, presenta mayor porcentaje de láminas $\beta$ que prpc. El cambio conformacional principal que distingue a prpsc de prpc es la ausencia de la $\alpha$-helice H1 de prpc que esta transformada en lamina $\beta$ en prpsc (20), esto afecta las propiedades fisicoquímicas de la proteina y su funcionalidad. Prpsc presenta escasa solubilidad en detergentes no ionicos, capacidad de formar agregados y resistencia parcial a la digestión con proteinasa K (20). La digestión de prpsc con dicha enzima produce un fragmento truncado en el extremo N-terminal denominado prp27-30 o prpres (por resistente in vitro). Además, en condiciones fisiológicas está favorecida la conformación rica en $\alpha$-helice de prpc siempre y cuando el gen PRP no este mutado. Muchas enfermedades causadas por priones se deben a mutaciones en el gen PRP que resultan en la síntesis de prpc con conformaciones alteradas ricas en laminas $\beta$ como la de prpsc. Se trata de autenticas moléculas de prpsc obtenidas en ausencia de infección por prpsc exogena. En la figura 2 se muestra la estructura de prpc basada en la información de (21).

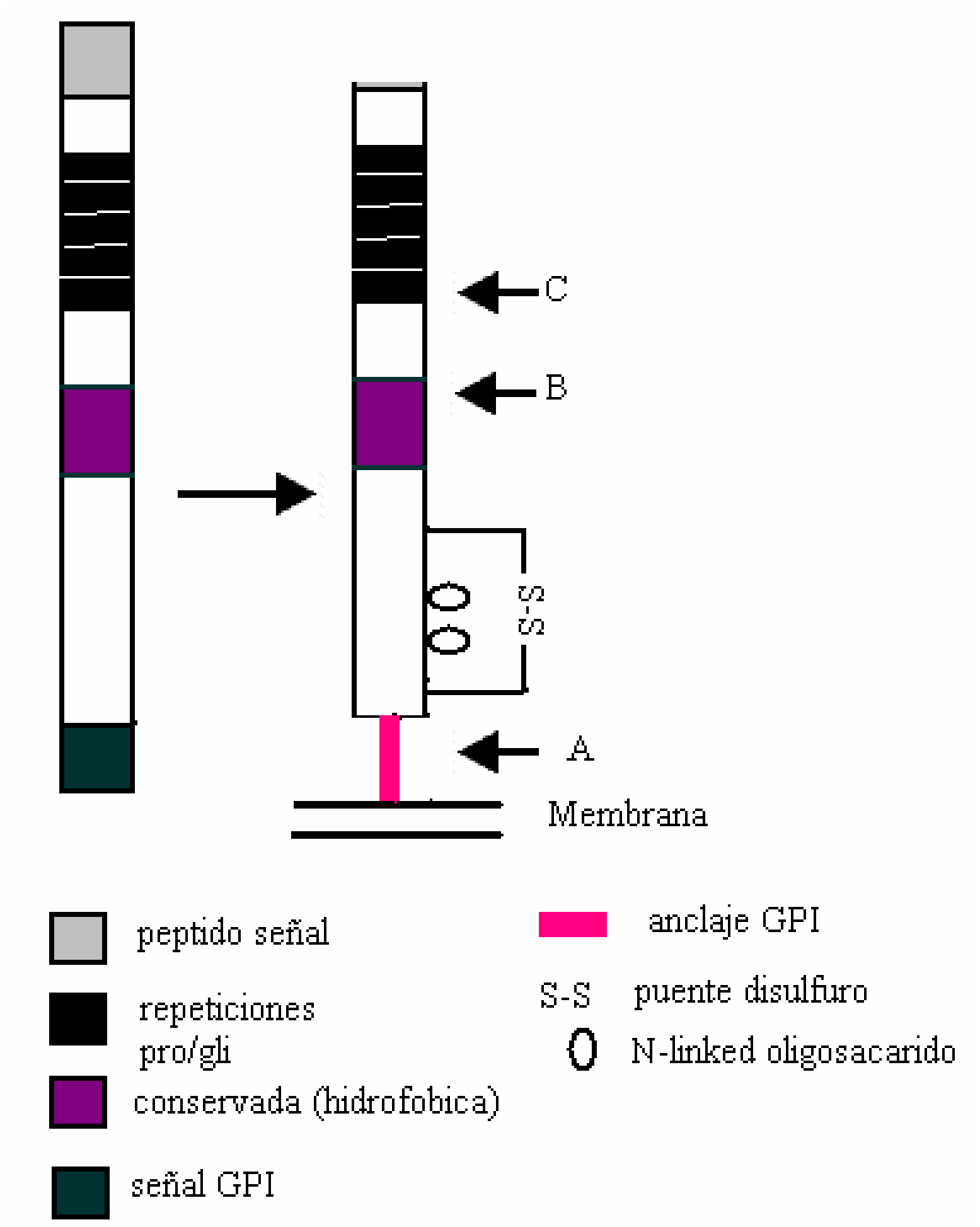

Fig 2. Estructura y procesamiento posttraduccional de prpc. En la parte superior se muestra la proteína recién sintetizada. Abajo se muestra la estructura de prpc madura. A y B son las posiciones de clivaje, C es la posición de clivaje que ocurre en la conformación prpsc cuando se la trata con proteinasa K, generando el fragmento prp37-30. B esta cercano al aa 110, C al aa 89 y A se encuentra en el GPI entre el motivo de diacilglicerol y el residuo de etanolamina.

## Funcion de prpc:

Esta proteína se localiza en la superficie celular (es una glicoproteína de membrana) esto sugiere que podría tener alguna función relacionada con la transducción de señales, uptake de ligandos, adhesión celular y reconocimiento celular. Prpc se vería imposibilitada de cumplir su función normal en presencia de prpsc, que le hace cambiar su conformación y esto sería responsable en parte de los síntomas que aparecen durante las encefalopatias espongiformes. Sin embargo, en experimentos con ratones en los que se hace knock out genico del gen PRP no aparece ningun cambio fenotípico, ni anormalidad visible (22), aunque otros autores encuentran efectos sobre el ritmo circadiano y el patron de sueño (23), anormalidades fisiológicas y estructurales en el hipocampo (24,25,26), pérdida de celulas de Purkinje en el cerebelo (27) y cambios en los patrones de aprendizaje y memoria (28,29). Aun asi, otros investigadores no encuentran estas anormalidades (22, 30, 31). Esto estaría relacionado con diferencias entre las líneas de ratón utilizadas que pueden tener diferencias en las regiones del gen PRP que se mutagenizan o se reemplazan, siendo mas o menos susceptibles a la ausencia de prpc funcional. Por otra parte, diferencias ambientales tambien pueden influir en el fenotipo. En ciertas condiciones, prpc podría ser necesaria, evidenciándose entonces una anormalidad fenotipica en los mutantes nulos, Por ejemplo un ratón mutante nulo para prpc expuesto a factores ambientales que aumenten su nivel de estrés oxidativo probablemente presente anormalidades en el fenotipo relacionadas con la ausencia de prpc ya que esta proteína interviene en la defensa antioxidante. Otra posibilidad es que simplemente, prpc sea una proteína redundante, es decir, tiene una funcion que también poseen otras proteínas, de modo tal que la ausencia de prpc esta siendo suplida total o parcialmente por otras proteinas, resultando en la ausencia de un cambio fenotipico visible. La realidad es que actualmente no existe una sola función de prpc de la cual haya podido determinarse inequívocamente que ocurre en las celulas in vivo salvo permitir la infección por parte de prpsc.

**Entre las funciones que se cree posee prpc se pueden destacar:**

El posible papel en el metabolismo del cobre y defensa antioxidante:

Como anticipé en el apartado anterior, prpc cumpliría funciones relacionadas con la defensa frente a los radicales libres. Esta defensa la ejercería mediante su capacidad de secuestrar iones cobre libres del medio celular (estos iones participan en reacciones de Fenton involucradas en la producción de especies oxidantes) y mediante su asociación con superoxido dismutasas (SODs) que son enzimas que participan de la defensa antioxidante convirtiendo el radical libre anión superóxido en peróxido de hidrógeno, a las cuales prpc les proporcionaria el cobre. Además existe evidencia de que la propia prpc tendría funcion enzimática de superoxido dismutasa.

Evidencia experimental a favor de dicha funcion de prpc:

Ratones nulos para PRP tienen menor contenido de cobre asociado a membrana plasmática y una menor actividad de superóxido dismutasas que requieren cobre como cofactor por lo que son mas propensos a extres oxidativo(32,33, 34, 35).
Ademas, tanto prpc recombinante como también péptidos derivados de prpc son capaces de unir iones cobre a traves de las repeticiones ricas en histidina presentes en la mitad N-terminal de la proteina. La región con repeticiones contiene 5 motivos en tandem de octapeptidos cuya forma general es P(H/G) GGGWGQ y está altamente conservada en las distintas especies de mamíferos, mientras que en la prpc aviar existe una region similar que consta de ocho repeticiones en tandem de octapéptidos (32).

El cobre parece ser importante para mantener la conformación de prpc ya que estabiliza las $\alpha$-helices de la proteína al unirse a esta en forma cooperativa. Esto se vio en estudios de espectroscopía que muestran que se unen cuatro iones cobre por molécula de prpc en las histidinas (37, 38). Además, como es necesario como cofactor de enzimas e interviene en la producción de radicales libres cuando esta en forma libre, es necesario su unión a proteínas (como por ejemplo prpc) y compartimentalización para minimizar sus efectos negativos.

El contenido de cobre en extractos de tejido cerebral enriquecidos en membrana de ratones knock out para el gen PRP es de 10 a 15 veces menor que en ratones control (que expresan prpc) mientras que no se observan variaciones en el contenido de otros metales. Esto sugiere que prpc es una proteína de union a cobre, importante para el metabolismo de este ión y podría controlar la actividad de otras proteínas asociadas a membrana que requieren cobre como cofactor (33). Además, mediante el empleo de cultivos primarios de células cerebelares expresando diferentes cantidades de prpc se encontro que las células con alta expresión de la misma tienen mayor resistencia al extres oxidativo en comparación a células prpc- (29). Esto sugiere que la capacidad de prpc de unir cobre es capaz de regular la actividad de enzimas involucradas en la defensa antioxidante, en particular superoxido dismutasas que requieren cobre como por ejemplo la SOD-1 y en consecuencia modular la respuesta al extres oxidativo. Si se sobreexpresa prpc en ratones wild type no se observa aumento de expresión de estas enzimas en ensayos de Northern, pero si su actividad. Los niveles de enzimas SOD son los mismos en ratones wild type, knock out para PRP y en los que sobrexpresan el gen, lo que varia es la actividad (29).

Ademas según algunos autores, prpc podria funcionar como un receptor para la endocitosis de cobre (30), mientras que para otros, la misma prpc tendría dicha actividad de superoxido dismutasa y uniria iones cobre para emplearlos como cofactor (32,33).

Interaccion con otras proteinas: supervivencia, diferenciación y transduccion de señales:

Aunque está demostrada la union de prpc a diversas proteínas, la importancia fisiológica de cada una de dichas interacciones aun no esta del todo clara y no siempre las interacciones encontradas en estudios in vitro se corresponden con lo que ocurre in vivo. Se requieren mas estudios para aclarar este punto.

Ciertas características estructurales de prpc permitirían su interacción con otras proteínas. Entre los sitios potenciales para las interacciones se encuentra una de las $\alpha$-helices ubicada en el centro de la molécula y el ancla de GPI (32).

Prpc es capaz de unirse a heparinas y compuestos similares a esta. La heparina es un polianión sulfatado similar a los glicosaminoglicanos. Los glicosaminoglicanos forman parte de la composición de las placas amiloides que se forman en presencia de prpsc. Aparentemente, las moleculas de heparina secuestran prpc impidiendo que se una a los glicosaminoglicanos por competencia. En un ensayo de union de prpc a los polianiones mediante la técnica de surface plasmon resonance muestra que la afinidad de prpc recombinantes por los polianiones esta correlacionada con una menor capacidad de interactuar con prpsc (36).

En un trabajo del grupo de Edenhofer (37) se utilizo el sistema de doble híbrido de levadura para buscar posibles proteínas que interactuan con prpc. En este estudio se encontro que una chaperona celular (Hsp60) es un ligando específico de prpc. El sitio de interacción fue mapeado entre los aa 180 y 210 de prpc. Posteriormente se encontro que la proteína Bip (otra chaperona) se asocia a prpc (32). Dicha proteína normalmente interactúa con proteínas mal plegadas o no ensambladas y media su translocación a los proteasomas donde son degradadas (38).

Prpc también se une a la proteína NCAM , una molécula de adhesión neuronal (39), al factor 2 relacionado a NF-E2 (Nrf2) que es un factor de transcripción (40), a Bcl-2 (una proteína antiapoptótica (32,41) y a la apolipoproteina E (ApoE), proteína de membrana que esta involucrada en la enfermedad de Alzheimer. Prpc se une al extremo N-terminal de ApoE a traves de su region N-terminal. Además esto ocurre en los hoyos revestidos de clatrina de la membrana plasmática. No se conoce que función cumple en la célula esta interacción y algunos autores incluso proponen que podria tener relación con la conversión de prpc a prpsc (42,43).

Otra proteína caracterizada como ligando de prpc es el precursor del receptor de laminina de 37Kda (44). Se encontró que existe interacción de este receptor y prpc tanto in vitro como in vivo y además está sobreexpresado en organos que acumulan prpsc (45)

Actualmente se cree que éste es el receptor de prpsc en las células de mamífero in vivo, aunque muy probablemente no sea el único.

Prpc también se une a la laminina, y el complejo laminina-prpc favorece la neuritogénesis de células en cultivo tratadas con factor de crecimiento neuronal (NGF)(46), por esto se piensa que prpc podría estar involucrada en el reconocimiento de ligandos y adhesién celular, mediante los cuales se disparan señales de proliferación y supervivencia en las celulas (32,47).

La laminina es una glicoproteina de 880 kD compuesta por dos polipéptidos grandes y uno mas corto dando un heterotrímero. Cumple funciones relacionadas con migración celular, adhesión celula-celula, apoptosis, diferenciación y proliferación celular, mediante su interacción con receptores de membrana. En el sistema nervioso central, la laminina participa de la diferenciación celular a traves de su interacción con integrinas y tambien de la migración de células y regeneración axonal. También evita la apoptosis de neuronas . Además, la laminina es degradada por la plasmina, que se produce a partir del plasminógeno por acción del activador de plasminógeno. La depleción de laminina en las neuronas es señal de apoptosis (47).

En ausencia de prpc la degradación de la laminina se encuentra aumentada, haciendo a las neuronas mas sensibles a la apoptosis. Prpc actúa como receptor de la laminina de alta afinidad y saturable. Prpc se une a la cadena $\gamma$ de la laminina (que es la mas conservada en todos los tipos de laminina) a traves de su extremo C-terminal (32) además puede interactuar con el plasminógeno (48).

PrPc y transdución de señales

A través de su localización en la membrana, prpc podría participar en vías de transducción de señal. Según algunos autores, la infección por priones afecta la función de canales de calcio. Por ejemplo, se vio que en células infectadas estimuladas con bradikinina (un estimulante de la función de los canales de calcio) presentaron una menor respuesta de los mismos en comparación a células control no infectadas (32).

Además, un fragmento de prpsc de 21 kD (el prp 106-126) es tóxico para cultivos primarios de células del hipocampo de rata. Este péptido es capaz de formar canales permeables a iones Ca y Na en membranas plasmáticas (32, 49). Tambien se observó que el péptido incrementa la concentración de calcio en células de microglia en cultivo obtenidas de ratones prp-/- lo cual activa vias de señalización (50). Dicho péptido permite la activación de las kinasas Lyn y Syk que participan de una cascada de transducción que resulta en la liberación de calcio intracelular y activación de la kinasa C y de una tirosin kinasa dependiente de calcio denominada PYK2. Luego ocurre la activación de MAP kinasas (ERK-1 y 2) (32).

Prpc dispara la cascada de señalización mediante el aumento del nivel de fosforilación de la tirosin kinasa fyn y la caveolina-1 es el factor intermediario entre prpc que esta del lado extracelular de la membrana y Fyn que es citoplasmática (32,43).

Por otra parte, la unión de prpc a la laminina aparentemente dispara la cascada de cAMP y la liberación de calcio (32).

Migración celular:

Prpc en celulas endoteliales que forman parte de la barrera hematoencefálica se acumula en las uniones celula-celula y participa de la transmigración de monocitos de los tejidos periféricos hacia el cerebro (51) posiblemente mediante reconocimiento específico de ciertas moléculas de la superficie de los monocitos.

## Biosintesis de prpc:

Prpc es sintetizada por ribosomas asociados al retículo endoplasmico. Al ser translocada al lumen del retículo, se cliva el peptido señal N-terminal, se añade un “core “ de oligosacaridos con union N en dos sitios (sitios A y B en la fig2), se forma un puente disulfuro y se adiciona el GPI. La proteína es transportada entonces en vesículas al golgi donde se procesa el oligosacarido, y se agrega acido sialico al GPI y tambien al oligosacarido (21).

Como resultado se obtiene la proteína madura prpc que es transportada a la membrana plasmática donde queda anclada por medio del GPI.
La cadena de oligosacaridos que se agrega inicialmente en el retículo es del tipo alta manosa y es sensible a la digestión con endoglicosidasa H. Esta cadena de oligosacarido es modificada en el golgi y queda como un oligosacarido complejo que contiene acido siálico y resulta resistente a la endoglicosidasa H (35).

El ancla de GPI se agrega en el retículo luego del clivaje del segmento hidrofóbico C- terminal y posee una porción central que es comun a otros GPIs, presentes en otras glicoproteínas, este esta compuesto de un residuo de etanolamina unido al aa C-terminal por una union amida, tres residuos de manosa, un residuo de glucosamina no acetilado y una molécula de fosfatidilinositol (PI) que se encuentra insertada en la membrana plasmática. El GPI de prpc (y de prpsc ) es poco comun porque posee acido siálico en la porción central, (52). La cadena de oligosacarido y el ancla de GPI de prpc es totalmente idéntica a la de prpsc, ambas isoformas son sintetizadas de la misma manera y sufren las mismas modificaciones posttraduccionales, existen diferencias en el oligosacarido entre las distintas cepas de prion (53).

La modificación experimental de la glicosilación de prpc altera el transporte de la misma. La mutación de los dos sitios consenso de glicosilación existentes en el gen PRP de raton o en la region N-terminal producen plegamiento anómalo de la proteína y su acumulación en un compartimiento proximal al golgi medio (54), Pero la glicosilación correcta no es totalmente necesaria para el transporte ya que la mutación del sitio consenso C-terminal o la síntesis de prpc en presencia de tunicamicina (un inhibidor de la glicosilacion) no impide que algunas moléculas lleguen a la superficie celular (54). Además, en ausencia de inhibidores es posible encontrar pequeñas cantidades de prpc no glicosilada en la membrana (35,54).
Por otra parte, las moléculas de prpc con alguno de los sitios de glicosilación mutados presentan características que las asemejan a prpsc, al igual que las que son sintetizadas en presencia de inhibidores de la glicosilacion (54).
En conclusión puede decirse que prpc tiene tendencia a adoptar la conformación de prpsc en condiciones fisiológicas durante su biosíntesis y transporte, pero las cadenas de glicano que posee ayudan a impedir este cambio conformacional. Además, esto es coherente con el hecho de que existen distintas cepas de priones y estas presentan diferentes patrones de glicosilación, que las distinguen entre sí (21). Parece que en la mayoria de los casos, lo que diferencia una cepa de otra es precisamente el patron de glicosilación.

Procesamiento postraduccional de prpc:

Una vez sintetizada, prpc es clivada en dos sitios distintos. Esto se vio a partir de experimentos con células transfectadas y en muestras de tejido cerebral y fluido cerebroespinal (35,55,56). Unos de los clivajes ocurre en el ancla de GPI liberando la proteína en el medio extracelular, pero no a todas, el resto queda anclada en la membrana.
Una fosfolipasa de membrana plasmática seria la responsable de catalizar el corte del GPI (57). El otro clivaje consiste en el corte proteolítico en un segmento de 16 aminoácidos muy conservado en las distintas especies que libera un péptido de prpc de 10kD que es un fragmento N-terminal de la proteína. Este clivaje ocurriría en dominios ricos en colesterol presentes en la membrana plasmática (56).
Se sabe además, que ambos tipos de procesamiento ocurren en forma relativamente lenta en comparación con la vida media de prpc, de modo que en el estado estacionario de la reacción, existe una población de prpc en distintos estadíos del procesamiento, es decir, existen todas las especies posibles: prpc intactas unidas a membrana por GPI, prpc libres en el espacio intercelular con y sin el corte en el dominio de 16 aminoácidos. No se sabe bien que funciones cumple cada producto proteico. Probablemente, la forma de 10 kD sea un ligando. Existen factores de crecimiento que se encuentran en forma de precursor unidos a la membrana plasmática y son liberados de la membrana por proteasas, dando el factor activo. Tal vez prpc se comporte de forma análoga. Otra posibilidad, es que prpc sea realmente un receptor de membrana (tal como gran parte de la evidencia experimental sugiere). Este receptor, podría en teoría existir en dos formas principales: anclado a membrana por el GPI o libre en el medio extracelular. (Además se pueden agregar como variantes de este modelo hipotético el estado clivado o no clivado en el dominio de 16 aminoácidos y la topología de la proteina, esto ultimo se explicara más adelante). En forma anclada, prpc se uniría a un ligando desconocido y participaría en alguna cascada de transducción de señales. La forma libre también uniría el ligando pero no podría transducir la señal. Esto es similar a lo que ocurre con receptores de factores de crecimiento y de citoquinas, que pueden existir en dos formas (soluble y anclada). La forma soluble es regulatoria, secuestra ligandos compitiendo con la forma anclada. Pareceria que en el caso de prpc ocurriria algo similar.

Localización de prpc en la celula y transporte:

En las moléculas de prpc ancladas a membrana, la unión a la misma consiste exclusivamente en el ancla de GPI (salvo para las formas transmembrana que son isoformas topológicas y seran tratadas posteriormente), por esta razón, prpc puede liberarse de la membrana por tratamiento con fosfolipasa C dependiente de fosfatidilinositol (PI-PLC), que es una enzima bacteriana que corta el GPI en el diacilglicerol (58).

En las neuronas, prpc es transportada por los axones hacia las terminaciones nerviosas (59), En estudios de inmunocitoquímica combinada con microscopía se vio que prpc se encuentra en terminaciones nerviosas de bulbo olfatorio y estructuras límbicas (21) por lo que podría tener alguna función relacionada con la sinapsis.

Mucho de lo que se sabe acerca del tráfico subcelular de prpc proviene de estudios con células transfectadas que expresan la proteína. En experimentos con prpc aviar (que es muy similar en estructura a la de mamíferos, comparten homologia), se vio que una vez sintetizada y anclada a la membrana plasmática no se queda alli sino que es endocitada, pasando al compartimiento endosomal. De hecho hay un reciclaje de prpc de la membrana hacia los endosomas(21,60). Esto es coherente con la posible funcion de prpc en el metabolismo del cobre, facilitaria la endocitosis del mismo mediada por receptor (la propia prpc actuaria como receptor del ión). Luego, los iones incorporados se disociarían de prpc debido al ph acido del endosoma y sería transferido a otras proteínas transportadoras de cobre. Dicho transporte de prpc se vio a partir de estudios en los que se utilizaron anticuerpos contra prpc marcados con un fluoróforo en cultivos celulares transfectados con el gen PRP. En estos estudios se siguió la marca del fluoróforo, que desaparecía de la membrana apareciendo en los endosomas(21).
Además, los iones cobre en concentración fisiológica estimulan en forma rápida y reversible la endocitosis de prpc (30).

<u>El papel de las Caveolas y vesiculas revestidas de clatrina</u>:

Muchos receptores celulares (como por ejemplo el receptor de LDL) se ubican en dominios de la membrana plasmática recubiertos de clatrina (chlatrin coated pits) La clatrina es es una proteína oligomérica de alto peso molecular formada por tres cadenas pesadas y tres cadenas livianas que se ensamblan formando una estructura llamada triskelion que polimeriza en la cara citosólica de la membrana plasmática. Esto produce la invaginación de la membrana que cuando es endocitada origina vesículas recubiertas de clatrina. Entre las estructuras celulares responsables de la endocitosis de prpc se encuentran estos dominios de membrana plasmática recubiertos de clatrina (21, 61).
Mediante microscopía electrónica se localizó prpc marcada con oro en las vesículas recubiertas de clatrina. Además, la incubación de células que expresan prpc en medio hipertónico con sucrosa (esto lleva a la disrupción de las uniones entre moléculas de clatrina) inhibe la endocitosis de prpc y en preparaciones purificadas de vesículas recubiertas de clatrina de tejido cerebral es posible encontrar prpc mediante anticuerpos específicos.
El segmento N-terminal de prpc es esencial para la endocitosis. Si se deleciona esta región, la internalización de la proteína disminuye considerablemente (62).
Se sabe que los receptores transmembrana que son internalizados por mecanismos dependientes de las vesículas recubiertas de clatrina en general poseen residuos de tirosina que se unen a la clatrina a traves de proteínas adaptadoras. Resulta extraño que prpc sea internalizada por dicho mecanismo ya que es una proteína anclada a membrana mediante GPI y este tipo de proteínas no posee dominios citoplasmáticos que puedan interactuar con la clatrina a traves de las proteínas adaptadoras (21). Algunos autores proponen que por esta razon prpc no puede ser endocitada en vesículas recubiertas de clatrina sino que en realidad lo haria mediante un mecanismo alternativo que involucra caveolas (63). Las caveolas también son dominios de membrana que presentan cierta invaginacion, pero estan recubiertos de una proteína denominada caveolina del lado citoplasmático en lugar de clatrina. Estas estructuras estan involucradas en señalización celular y uptake de ligandos. Son abundantes en fibroblastos, celulas musculares lisas y células endoteliales. Sin embargo, células neuronales que expresan prpc y que no expresan caveolina son capaces de internalizar prpc esto sugiere que o bien prpc es endocitada a traves de la via endocitica clásica dependiente de clatrina a pesar de ser una proteina con GPI o lo hace por medio de otro mecanismo en dichas células. Aparentemente, en las neuronas prpc se ubica en microdominios de membrana que poseen otras proteínas y cierta composición de lípidos particular. Estas regiones se denominan lipid Rafts (balsas lipídicas) y actualmente el modelo aceptado para la endocitosis de prpc involucra a estos dominios mas que a la clatrina (21). Por otra parte, el mecanismo via clatrina podria ser correcto si prpc interactuara en la membrana con alguna proteína receptora en los hoyos recubiertos de clatrina. Esta proteína aportaría el dominio citosólico que se une a las proteínas adaptadoras que esta ausente en prpc y ademas poseería un dominio extracelular capaz de unirse a la porcion N-terminal de prpc. La unión de iones cobre a prpc estimularía la unión de ésta a su receptor y en consecuencia se favorecería la endocitosis del complejo cobre-prpc unido al receptor. Dicho receptor aun no fue identificado inequívocamente, mas bien se encontraron varios posibles candidatos, por ejemplo laminina, NCAMs,etc. Una aproximación experimental para encontrar el receptor es la búsqueda de sitios de union en prpc para proteinas conocidas. En un estudio de este tipo realizado con proteinas de fusion conteniendo el fragmento N-terminal de prpc se encontró que dicha región posee sitios de unión para glicosaminoglicanos (GAGs), en especial heparan sulfato (21).

Prpc se asocia a lipid rafts ricos en colesterol y glicoesfingolípidos a traves de la union de su ancla de GPI con lípidos saturados presentes en estos dominios de membrana y también mediante la interacción de su extremo N-terminal con proteínas presentes en los rafts como por ejemplo la caveolina-1. Dichos dominios lipídicos son resistentes a detergentes y presentan diversas proteínas ancladas por GPI (21).

En algunas células, prpc es endocitada a traves de caveolas, pero en las neuronas, una vez que une cobre en las repeticiones de octapéptidos de la region N-terminal, se transloca fuera de los rafts y va a regiones de membrana que se solubilizan cuando son tratadas con detergentes desde donde accede a los hoyos recubiertos de clatrina, donde es endocitada.
La region N-terminal de prpc es importante para la union de la proteína adaptadora ubicada en las regiones recubiertas de clatrina.
La asociación de prpc a los rafts ocurre a traves de diversas moléculas involucradas en la señalización celular, entre las que puede destacarse la caveolina-1 y las tirosinkinasas Fyn y Scr. El agrupamiento de prpc en los rafts dispara procesos relacionados con la tranducción de señales, entre ellos, el reclutamiento de moléculas de adhesión celular que promueven el crecimiento de las células neuronales. Ademas, los lipid rafts funcionarían como un microambiente favorable para la conversión de prpc en la forma patológica prpsc (64).

En cuanto a la localización de prpsc en las células, en estudios con técnicas de inmunofluorescencia en celulas N2A infectadas con prpsc se encontro que parte de las moleculas de prpsc se encuentran en el interior celular, donde colocalizan con marcadores de aparato de Golgi (65), ademas también se encontró que colocaliza con proteínas marcadoras de endosomas y lisosomas (66,67) y también aparecen en la membrana plasmática, donde son observables mediante microscopía electrónica (66). En base a estos resultados, es evidente que prpsc sigue las mismas vias de localización celular que prpc, lo cual es de esperarse ya que al igual que prpc posee ancla de GPI, lo cual restringe su localización a la membrana plasmática y ademas esto le permite acceder a la via endocítica, con lo cual va sufrir transporte hacia endosomas y lisosomas. Por otra parte, las moléculas de prpsc sintetizadas de novo en las células tienen la señal de translocación al retículo y de adición de oligosacáridos que posee prpc. Se sintetizan y procesan de la misma manera descripta para prpc.

Topologia de prpc en la membrana plasmática

Normalmente, la topología de una proteína (esto es, su ubicación y orientación en el espacio) es única y está determinada por elementos topogénicos en su secuencia de aminoácidos (68). De esta manera, todas las moléculas de una misma proteína no solo se sintetizan de la misma manera y tienen el mismo procesamiento sino que se ubican de la misma forma en el espacio, por ejemplo todas con el N-terminal del lado del citosol.

En el caso de prpc, esto no se cumple. Esta proteína posee tres elementos de secuencia con funcion topogenica: el peptido señal N-terminal que la destina al reticulo, un segmento de a-helice hidrofóbico como el que poseen los dominios transmembrana de otras proteínas, y la secuencia señal para adicion del GPI en el extremo C-terminal (68).

Por alguna razon desconocida, los elementos topogénicos fallan en determinar una unica topologia para prpc, de esta manera, en condiciones normales, la población de moleculas de prpc se divide en tres subpoblaciones cada una con una topologia particular y se las puede denominar isoformas topológicas (68).

Por lo general cuando una proteína tiene más de un elemento topogénico, predomina uno de ellos, se dice que hay secuencias señal mas fuertes que otras, las secuencias llamadas débiles, por ejemplo pueden ser aquellas que estan ubicadas en regiones de la proteína cuya conformación interfiere con que sean reconocidos por las proteínas que participan del transporte y translocación. En el caso de prpc dicha secuencia debil sería la región hidrofóbica, predominando la secuencia del GPI. La proteína del prion es especial en cuanto a su capacidad de variar su conformación, probablemente, la region hidrofóbica que está funcionando como señal transmembrana débil, se pliega de una forma que esconde los aa hidrofóbicos del medio y resulta bastante estable en esa forma, (pero no tanto como para que toda la población de prpc se pliegue asi) con lo cual no se inserta en la membrana y se le añade el GPI.

Las isoformas topológicas de prpc son:

Una forma de secreción denominada secprpc, esta es la isoforma mayoritaria y se trata simplemente de la prpc clásica anclada por GPI del lado extracelular de la membrana plasmática descripta en el apartado de biosíntesis. Esta isoforma es transportada en vesículas de secreción a la superficie exterior celular donde queda anclada a través del GPI en dominios de membrana caveolas o invaginaciones recubiertas de clatrina según se detallo anteriormente. Una vez allí, algunas moléculas son liberadas al espacio extracelular por escisión del GPI mientras que la mayoría es internalizada en el compartimento endocítico y realiza ciclos entre este compartimiento y la membrana (69).

Una forma transmembrana con el extremo N-terminal del lado del lumen del retículo (igual que secprpc) y el C-terminal del lado citosolico denominada NTMprpc por prpc transmembrana (TM) con N-terminal luminal (N).
Otra forma transmembrana con el C-terminal del lado del lumen del retículo y el N-terminal del lado citosólico (denominada CTMprpc análogamente a NTMprpc)

La isoforma secprc se considera normal y no produce neurodegeneracion. En cambio, CTMprpc es una proteina transmembrana que se acumula en el Golgi y se comporta como un intermediario neurotoxico (69).

Existen mutaciones en el gen PRP que favorecen la capacidad de prpc de adoptar la forma CTMprpc. Dichas mutaciones se asocian a encefalopatias espongiformes en humanos y ratones. Por ejemplo una mutación denominada prp A117V produce la enfermedad GSS en humanos (detallada mas adelante en el apartado de enfermedades) y presenta aumento de la concentración de CTMprpc (19,68,69,70). Dicha isoforma es la menos estable de las tres y presentaría una flexibilidad conformacional mayor con lo cual aumenta la probabilidad de convertirse en prpsc en forma espontanea. Esto sería al menos en parte responsable de la aparición de encefalopatia espongiforme en individuos con mutaciones en PRP. Otras mutaciones que afecten otras regiones de la proteína también pueden resultar en la sintesis de prpc con conformación alterada patológica. Asimismo, variaciones en el ambiente celular debidas a diversos motivos pueden estabilizar en mayor o menor grado alguna de las formas topológicas de prpc favoreciendo según el caso la aparición de una encefalopatia espongiforme esporádica en ausencia de mutaciones en el gen.

El mecanismo responsable de originar las isoformas topológicas es desconocido (70). Según un modelo (70), el primer paso es la translocación cotraduccional del péptido de prpc que se esta sintetizando mediante el reconocimiento del peptido señal N-terminal por parte de proteínas asociadas al translocón (canal del retículo por donde ingresan proteinas cotraduccionalmente).
Una vez dentro del translocón, la población inicial de prpc que era homogénea hasta ese momento es particionada en dos subpoblaciones (una formada por prpc con extremo N-terminal del lado del lumen del retículo y otra con el C-terminal del lado del lumen. La síntesis prosigue y entonces, cuando aparece la porción hidrofóbica de prpc se produce un segundo evento de partición que separa la subpoblación de prpc con N-terminal luminal en formas transmembrana (NTMprpc) y forma secprpc. Esto ocurriría porque durante la translocación, hay moléculas que acceden al translocón en distintos estadíos de síntesis, algunas se translocan cuando solo tienen el peptido señal N-terminal entonces empiezan a plegarse con ayuda de als chaperonas del retículo un poco antes de que se sintetice la region hidrofobica. De esta manera, cuando aparece dicha region hidrofóbica, el grado de plegamiento es tal que queda desfavorecida la inserción a la membrana entonces se obtienen la forma secprpc.

Si en cambio la proteína se empieza a translocar cuando se esta sintetizando la región hidrofóbica o un poco antes de eso, las chaperonas no tienen tiempo de plegar tanto al peptido de prpc y se establece una competencia entre el segmento hidrofobico y la region N-terminal de la proteína. Si gana la región hidrofóbica, se obtiene la forma transmembrna NTMprpc, de lo contrario se obtiene secprpc pues la porcion N-terminal no favorece la inserción en la membrana.

En el caso de la forma CTMprpc, como tiene el extremo C-terminal del lado del lumen del retículo y el N-terminal del lado contrario, el segmento hidrofobico no puede competir con el N-terminal, entonces gana y se inserta en la membrana, obteniéndose la forma transmembrana CTMprpc.

**Modelos para la propagación de prpsc**

El proceso de propagación del prion es un proceso exponencial que se inicia con la interacción de la prpsc exógena con prpc o con alguna forma de prpc parcialmente desnaturalizada a traves del reconocimiento de la region 96-167 por parte de prpsc en forma dependiente de la identidad de secuencia. Una vez que la interacción ocurre, la formación de mas moléculas de prpsc se vuelve exponencial y la velocidad conversión es directamente proporcional al nivel de expresión de prpc (32). La presencia de prpc es absolutamente necesaria para que prpsc se propague. Ratones knock out para el gen prp son incapaces de desarrollar enfermedad cuando son inoculados con prpsc (20).

Prpsc interactua con prpc durante la reaccion de replicacion de prpsc y esta interacción es más eficiente cuando las dos isoformas comparten la misma secuencia de aminoácidos. Esta es la base de la barrera de especie que existe para la transmisión de priones de una especie a otra. La secuencia de aa de prpc varía en algunas posiciones de una especie a otra y la reacción de conversión es más eficiente cuanto mayor sea la similitud entre las dos isoformas interactuantes (8,18).

Existen dos modelos que explican como ocurre el cambio conformacional de prpc en presencia de prpsc, que se muestran en la figura 5 (32, 68):

Modelo del replegamiento de prpc

Según este modelo, la aparición de una molécula de prpsc cataliza la conversion de las moléculas de prpc en prpsc, para esto, prpc se desnaturaliza parcialmente y lugo vuelve a plegarse bajo la influencia de prpsc, tomando la conformación de esta. Asi, prpsc actuaria como molde en esta reaccion, posiblemente asistida por otras proteinas, especialmente chaperonas. Las moléculas de prpsc, producto de la reacción a su vez pueden actuar como molde para convertir mas moleculas de prpc.

El cambio conformacional es la etapa limitante del proceso y en ausencia de prpsc no se produce debido a la alta barrera de energia de activación. se requiere para bajar la energia de activacion necesaria para producir el cambio conformacional de prpc La reacción de conversión según este modelo ocurre mediante un mecanismo similar al de las reacciones catalizadas enzimáticamente:

1) Prpc se comporta como sustrato y prpsc es el producto de la reacción.
2) La velocidad de la reacción depende de la concentración del sustrato.
3) Prpsc exogena funciona como un efector alostérico que regula la conversión de prpc en prpsc.
4) la presencia de análogos del sustrato, por ejemplo moléculas de prpc distintas provenientes de distinta especie, puede inhibir la reacción comportándose como inhibidores competitivos.

La reacción entre una molécula de prpsc con una de prpc da como producto dos moléculas de prpsc. Cada una se une a una molécula de prpc con lo cual se obtienen cuatro moléculas de prpsc (dos por reaccion) y asi sucesivamente. De esta forma la cantidad de prpsc crece exponencialmente.

<u>Modelo de la “semilla</u>” (seeding model):

Según este modelo, prpc y prpsc se encuentran en equilibrio, con la conformación prpc favorecida. La forma patológica prpsc se estabiliza mediante su agregación. La velocidad de asociación de las moleculas de prpsc en agregados es muy lenta y esta desfavorecida, pero una vez que se forma un agregado o núcleo este actua como semilla en la polimerización con lo cual la velocidad de union de mas moleculas de prpsc aumenta y entonces el sistema es perturbado con lo cual se desplaza el equilibrio hacia la formación de mas prpsc. Este modelo en principio no explica el aumento exponencial de la concentración de prpsc que se observa durante una infección, pero si se supone que los agregados formados una vez que llegan a cierto tamaño se rompen originando nuevos núcleos, entonces el aumento de prpsc puede ser exponencial (68).

Ambos modelos justifican las tres variantes que existen de las enfermedades producidas por priones (infecciosas, familiares y esporádicas) (32).

Las patologías infecciosas serían el resultado de la presencia de Prpsc exógena, que cataliza la conversión de prpc según el modelo de replegamiento o altera el equilibrio a favor de si misma según el modelo de polimerización nucleada.

En las patologías hereditarias ocurre una desestabilización de la estructura de prpc o una estabilización de la estructura de prpsc favoreciendo la población del estado patológico. Por último, las enfermedades esporádicas, podrían surgir por alteraciones metabólicas o bien mutaciones espontáneas que conlleven la formación de prpsc.

Las chaperonas celulares pueden intervenir en la conversión de prpc en prpsc tanto en forma positiva como negativa:

En diversos estudios se observo que ciertas chaperonas son capaces de favorecer la conversión de prpc en prpsc en presencia de prpsc exogena.

En ausencia de chaperonas, la conversion de prpc en prpsc es lenta e ineficiente. In vitro, las chaperonas reconocen y se unen moléculas de prpc nativas y desnaturalizadas por ejemplo mediante tratamiento con medio acido, alterando su conformación, esto facilita el reconocimiento de prpc por parte de prpsc, con lo cual se asocian y la reacción de conversión ocurre con alta eficiencia. De esta manera, las chaperonas facilitan el primer paso de la reacción que es la interacción entre ambas conformaciones del prion (32, 71).

En estudios in vitro con prpc marcada radiactivamente, se observó que las chaperonas celulares aumentan la eficiencia de dicho paso lento de reacción mediante la union a prpc, cuando la chaperona se une a prpc, le produce un cambio conformacional que origina un intermediario de prpc que es proteasa sensible pero puede recuperarse por centrifugación en forma de agregados en el pellet. El intermediario de prpc es reconocido por prpsc mas que la prpc nativa con lo cual se favorece su conversión en prpsc.

En el segundo paso de la reacción, prpsc produce un cambio conformacional en el intermediario de prpc que entonces se vuelve proteasa resistente y adquiere la conformación de prpsc.

Por otra parte, otras chaperonas interfieren con la reacción mediante la union a prpc. En este caso, la unión de la chaperona estabiliza una conformación de prpc que no resulta favorable a la interacción con prpsc. Esto constituye una especie de sistema de defensa de la celula infectada con prpsc que desfavorece la propagación de esta, se cree que este es uno de los motivos por los cuales las enfermades de origen prionico presentan tiempos de incubacion tan prolongados. Entre las chaperonas que interfienen con la conversión se encuentra Bip, en cambio, las chaperonas Hsp60 y Hsp104 facilitan la reacción (38).

La barrera de especies

Existe especificidad de especies para la transmisión de priones, lo que se conoce como la barrera de especie, propuesta por Pattison (17, 18,32,72), que se manifiesta como un periodo de incubación más prolongado que lo esperado al pasar de una especie a otra. Este efecto se observa aún entre especies muy cercanas, esto fue demostrado con ratones y hamsters (18, 73).

La transmisión de priones entre distintas especies es un proceso estocástico. En el caso de ocurrir, este proceso es muy poco eficaz y sucede con una prolongación del tiempo de incubación que tras pases subsiguientes, o adaptación, se acorta y estabiliza y la transmisión deja de ser un proceso probabilístico. Los priones sintetizados de novo reflejan la secuencia del gen PRP del huésped. En los primeros estudios al respecto se puso de manifiesto que la barrera de especie era la manifestación de las restricciones de secuencia de un proceso de reconocimiento molecular. Así, la infección ocurre a través de un complejo prpc-prpsc, cuya formación está gobernada por el grado de identidad de secuencia entre la proteína endógena y la exógena. La identidad en el segmento 96-167 es necesaria pero insuficiente la interacción de prpc con prpsc es más eficaz cuanto mayor es la identidad de secuencia en la región 96-167. Se requiere además la presencia de una proteina adicional que reconoce la region C-terminal de prpc. Dicha proteina es una chaperona, originalmente denominada proteina X. La interacción de prpc con la chaperona es más eficiente cuando ambas son de la misma especie (32).

La existencia de distintas cepas de priones:

Una de las características de los priones es su multiplicidad, es decir, la existencia de multiples cepas, (tambien denominadas inóculos o strains) similar a lo que ocurre con los virus y bacterias. Las cepas se diferencian en el tiempo de incubación y en los patrones de lesion en el sistema nervioso cuando son inoculados en lineas de raton o hamster estandarizadas (32, 74). En esto se basa un ensayo en animales de laboratorio de valor diagnostico y confirmatorio de TSEs (ver seccion diagnostico).
Todas las cepas de prpsc tienen la misma secuencia de aminoácidos pero se diferencian en el patron de glicosilacion y en el tamaño de los fragmentos que se originan cuando son incubadas con proteasas, esto se debe a que presentan variaciones en su conformación que hacen que ciertas regiones de la proteina esten más o menos expuestos a la accion de las proteasas. Las características propias de cada cepa de prpsc son transferibles a prpc durante la reaccion de cambio conformacional, pues prpc adquiere la conformacion exacta de prpsc. Ademas, para que la conversión ocurra, la secuencia aminoacídica de prpc no solo debe ser lo mas similar posible a la de prpsc según el postulado de la barrera de especie sino que el patron de glicosilación también debe coincidir (32, 74).

El hecho de que las distintas cepas de priones produzcan distintos patrones de lesiones cerebrales permite distinguirlas y tiene valor diagnostico. Mediante análisis computarizado de imágenes de cortes de tejido se puede distinguir cada cepa, pues el patron de lesiones funciona como un sello de identidad util en el diagnostico (18).

Metodos de inactivacion de priones:

Muchas de las características biológicas de los priones son similares a aquellas correspondientes a los virus. Sin embargo, se distinguen de los mismos debido a la marcada resistencia que presentan frente a los agentes inactivantes como el calor y las radiaciones. Dicha resistencia frente a la mayoría de los agentes inactivantes que se utilizan contra virus y bacterias plantea un grave problema a la hora de decidir qué procedimientos pueden asegurar la destrucción de la infectividad (18).

Características físicoquímicas y biológicas de los priones prpsc:

Propiedades físico - químicas:

Filtrables (tamaño de poro: 0.22 µm)

No presentan estructuras de partículas (como los virus) definidas al microscopio electrónico

Resistentes a: autoclavado, formol, radiaciones, etanol, agua oxigenada, nucleasas, parcialmente resistentes a proteasas.

Propiedades biológicas:

Prolongado período de incubación

No inducen respuesta inmune en el hospedador infectado

Existencia de cepas (“strains”)

Composición proteica exclusivamente: proteína del prión (PrP)

Presentan barrera de especie

Los procedimientos que logran reducir el título de la infectividad a valores razonables incluyen: hipoclorito de sodio 10% (1hora), hidróxido de sodio 2 M, permanganato de potasio 0,002 M, fenol 90%, autoclavado a 134°C, 30 min, cloroformo, éter, acetona y urea 6 M (9,18).

Los procedimientos mencionados aseguran la disminución de la infectividad de materiales contaminados por priones, pero aun asi existen evidencias de que materiales sometidos a una temperatura de 600°C, a la cual la materia orgánica ya se ha descompuest, conservan restos de infectividad. Este hecho podría explicarse considerando que debido a la carbonización que sufre la materia orgánica a esa temperatura, se podrían formar “moldes” de carbón sobre las moléculas de PrPSc y esos moldes inorgánicos serían los que inducirían a PrPc a plegarse de manera incorrecta (18).

## Aproximaciones experimentales para el estudio de priones:

En un principio, los priones se empezaron a estudiar mediante inoculaciones de material infectado en ovinos, caprinos y monos. El empleo de estos animales tenia desventajas obvias que impedían el rápido avance en las investigaciones por lo que fue necesario obtener ratones y hamsters transgenicos con los cual surgieron los primeros modelos experimentales de encefalopatias espongiformes. Estos modelos incluyen animales sin el gen que codifica la proteina del prion, animales quiméricos y otros que expresan fracciones de Prnp que originan versiones de Prpc truncadas en alguno de sus dominios.

En distintas investigaciones se obseróo que los ratones desprovistos del gen PRP (y por lo tanto sin Prpc) no podian contraer enfermedades de origen prionico y generaban anticuerpos contra los priones cuando estos eran inoculados. Ademas, como estos ratones eran viables se llego a la conclusión de que Prpc no era una proteína esencial , tal vez se trataba de un producto génico redundante. Aún asi, diversos estudios demostraron que si bien los animales carentes de Prpc aparentan ser normales, pueden presentar fallas en su ciclo circadiano, algunos cambios leves de comportamiento, problemas de aprendizaje y muy raramente, alteraciones en su desarrollo. Sin embargo, esto podría depender de la cepa de ratón y de condiciones experimentales mas que de la ausencia de prpc .
En cuanto a la generación de anticuerpos anti priones por los ratones sin el gen PRP (denominados ratones Tg 0/0, donde Tg indica transgenico) se debe a que al estar ausente Prpc, no se genera tolerancia inmunológica frente a esta proteína en etapas tempranas del desarrollo, entonces las células del sistema inmune al encontrarse con los priones inoculados no los reconoce como propios del organismo y generan una respuesta inmune típica (en cambio, si el animal expresa Prpc, la tolerancia se establece normalmente y el sistema inmune frente a una infección por priones exogenos es incapaz de diferenciarlos de la isoforma celular Prpc) (21).

## Modelos de raton empleados en el estudio de los priones: (74)

(Mo)Prn-P +/+ : Ratón (Mo) con ambos genes propios. Los animales infectados con PrPsc enferman tras incubación corta.

(Mo)Prn-P o/+ : Ratón con ablación de un gen. Los animales infectados enferman tras larga incubación.

(Mo)Prn-P o/o : Ratón desprovisto de ambos genes. Los animales infectados no enferman de EE, pero presentan ataxia a partir de los 490 días de edad

Ratón Tg Prn-P Hu/Mo : Ratón transgénico con una quimera del gen Prn-P humano/ratón.

Tg(MHu2M)Prn-P :Es una quimera formada por 2 pares de genes: un par ratón/humano y otro par ratón/ratón. Esta quimera es altamente susceptible a priones humanos y presenta un período corto de incubación.

Ratón Tg Prn-P SHa/Mo :Quimera hámster sirio/ratón. Inoculada con priones hámster sirios (SHa) produce muchas placas de amiloide, pero no si se inocula con priones de ratón. Con estos Tg también se han obtenido priones PrPsc quiméricos SHa/Mo.

Ratón Tg(GSS MoPrP : Quimera de gen humano con mutación que produce la enfermedad de GSS y gen normal de ratón; produjo espontáneamente procesos neuro-degenerativos.

Ratón Tg múltiple : Quimera con 7 juegos de Prn-P de SHa. En homocigosis el ratón enferma espontáneamente después de 400 a 600 días, y en hemicigosis después de 650 días.

(Nota sobre la nomenclatura empleada: Mo indica gen de raton (mouse), Ha indica gen de hamster y Hu indica que la copia del gen es humana).

Para definir el mecanismo por el cual prpsc convierte a prpc en copias de si misma se pueden utilizar diversos métodos experimentales, entre los que se destacan:

Análisis estructurales de las isoformas prpc y prpsc mediante técnicas de espectroscopia NMR, cristalografia de rayos x , (8,72,75).

Ensayos de unión mediante la técnica de surface plasmon resonance.

Sistemas in vitro con péptidos del prion purificados y prpc, con los cuales se realiza la conversión de prpc en prpsc in vitro. En estos modelos se varian parámetros tales como por ejemplo la secuencia del péptido y sirvieron para encontrar las regiones de prpsc responsables del cambio conformacional (21).

También se utilizan mucho técnicas tipicas de biología celular como por ejemplo técnicas de marcado metabólico, cultivos celulares y fraccionamiento subcelular entre otras técnicas para determinar cual es la vía de biosíntesis de prpc, ubicación dentro de la celula, procesamiento posttraduccional. Este abordaje tiene la ventaja de que permite estudiar los priones dentro del contexto fisiológico normal, en presencia de proteína nativa y los cofactores necesarios para la reaccion de conversión.

**Líneas celulares empleadas comunmente en la investigacion de los priones:**

Se desarrollaron diversas líneas celulares derivadas de neuronas para el estudio de los priones. Entre estas se destaca la linea N2A derivada de células de neuroblastoma de ratón, que es una de las más utilizadas (21). Otras lineas son: la PC12 de feocromocitoma de rata (76), HaB de hamster (células inmortalizadas espontáneamente), celulas GT1 (neuronas hipotalamicas ) (77)
Todas estas lineas producen continuamente prpsc luego de ser infectadas y si bien en general no muestran signos patológicos, las células GT1 infectadas pueden entrar en apoptosis (77) y las N2A infectadas presentan alteraciones en la fluidez de la membrana y en la respuesta a bradikinina siendo por lo demas viables (21, 78).
Actualmente se dispone de líneas celulares que expresan genes PRP con cualquiera de las mutaciones conocidas que producen enfermedad por priones en humanos, ratones y hamsters.

**Enfermedades causadas por priones: Las encefalopatias espongiformes**

Las encefalopatias espongiformes transmisibles (TSEs), son enfermedades neurodegenerativas fatales, con periodo de incubación lento, pueden tardar años o decadas en manifestar sintomas, y actualmente no tienen tratamiento. Se producen debido a la acumulación de priones prp (específicamente la conformación prpsc de dicho prion) en el sistema nervioso. Esto desencadena la muerte neuronal por apoptosis. Este tipo de enfermedad incluye la enfermedad de Creutzfeldt- Jacob (CJD), el síndrome de Gerstmann-Strausler (GSS), el insomnio fatal (FFI) y el kuru entre las que afectan a los humanos, pero se conocen muchas más que son propias de otras especies animales (Tabla 1), como por ejemplo la enfermedad del cansancio crónico de los cervidos, el scrapie del ganado ovino y la renombrada encefalopatia espongiforme bovina (BSE), tambien conocida como el mal de la vaca loca. Esta última tiene una variante surgida recientemente que es capaz de infectar al ser humano dando los que se conoce como nueva variante de CJD y se cree que la via de transmisión es a traves de carne vacuna contaminada con el prion de la BSE.

El conocimiento del origen infeccioso de las TSEs tiene larga data: En 1939 se logro transmitir el scrapie a cabras mediante inyección intraocular (74,79). En 1950, Carleton Gajdusek describio el kuru, que es endémico de una tribu de Nueva Guinea y se transmite entre los miembros de la misma a traves del canibalismo ritual. Posteriormente el veterinario William Hadlow noto las similitud neuropatologica entre el kuru y el scrapie. En trabajos siguientes se demostro que el kuru y el CJD son transmisibles a primates. Estos trabajos despejaron toda duda acerca de la naturaleza infecciosa de estas enfermedades (21, 74).

A principios de los años 80, Stanley Prusiner propuso la hipótesis del prion (72), esta establece que el agente infeccioso responsable de las encefalopatias espongiformes esta compuesto exclusivamente por un único tipo de proteína a la cual denominó prion y actualmente se la conoce en la literatura como prpsc (por proteina del prion del scrapie) En trabajos posteriores se demostró que prpsc es una versión alterada de una proteina celular normal, codificada por el hospedador, prpc (por proteina del prion celular) y que ademas es una proteina altamente conservada que se encuentra en todas las especies analizadas hasta la actualidad.

Actualmente se sabe que prpsc propaga su conformación entre las moléculas de prpc con lo cual todas quedan con la misma conformación que prpsc. La reaccion es autocatalitica. Asi, los priones presentan un novedoso sistema de propagación que contradice el dogma central de la biología pues se trata de transmisión de información a traves de los cambios conformacionales de una proteina y no a traves de un acido nucleico.

Por otra parte, las enfermedades producidas por los priones son únicas porque pueden ser tanto de origen genetico (debido a mutaciones en el gen PRP en la linea germinal ), esporádico (por mutación somatica al azar no determinada) o transmisible. Son las únicas enfermedades conocidas que pueden ser a la vez geneticas (familiar) e infecciosas (80,81).
Dado que el agente que las produce tiene la misma secuencia de aa que prpc, en estas enfermedades no aparecen procesos inflamatorios ni respuesta del sistema inmune (82).

En terminos generales, los síntomas que presentan son demencia, disfunción motora, ataxia cerebelar, disautonomia y en el caso particular del FFI aparece insomnio progresivo. Dichos síntomas aparecen luego de un periodo largo de años a decadas luego de la exposición inicial al patógeno. En el caso del kuru, FFI y CJD aparecen a edades avanzadas, en general a partir de los 50 años de edad, luego de décadas de incubación, aunque vCJD se manifiesta a edades trempranas, en la juventud. Por otra parte, GSS solo tiene unos 10-12 años de incubación antes de que los síntomas comiencen a manifestarse (80).
Por otra parte, en cortes de tejido nervioso se observa espongiosis junto con atrofia del tejido nervioso por perdida neuronal asi como también deposición de placas de tipo amiloide, éstas son más prominentes en el GSS y el Kuru y se asemejan a la amiloidosis producida en la enfermedad de Alzheimer, que también es producida por una proteína con plegamiento anomalo (80).

Existe evidencia de que vCJD se originó a partir de la transmisión del prion de la BSE a humanos a partir de carne contaminada. Además, el prion de la BSE seria de origen ovino, derivado del prion causante del scrapie del ganado ovino, este habría cruzado la barrera de especie alrededor de 1986 cuando se comenzo a alimentar al ganado con harinas de hueso y carne proveniente de ovinos infectados de scrapie. Hacia 1988 la epidemia resultante de BSE fue suprimida cuando se interrumpio esta practica, pero en 1996 hubo otra epidemia por el mismo motivo y a demas aparecieron los primeros casos en humanos (vCJD) (66).
Ademas se conocen otra vía de trasmisión de CJD aparte de la oral que es la via iatrogénica a traves de inyecciones de hormona de crecimiento obtenida de pituitaria, transplantes de cornea, intrumentos quirúrgicos, etc, pues los priones son bastante resistentes a los métodos de esterilización (83,84).

Se estima que el 10 % de los casos de CJD y todos los de GSS y FFI se deben a mutaciones en la linea germinal en el gen PRP en el cromosoma 20 (17). Dichas mutaciones favorecen el cambio espontaneo de conformación de prpc a la forma patogénica prpsc sin necesidad de que exista infección con prpsc exogena. Mutaciones puntuales en el extremo C-terminal de prpc están asociadas a la aparicion de FFI, CJD y GSS. Por otra parte, mutaciones de inserción en el extremo N-terminal estan asociadas a la aparicion de CJD y ademas consisten en la repetición de un octapéptido presente en la proteína
(1 a 9 copias del mismo). Dicho octapéptido es normal que este repetido solo 5 veces. Existe un polimorfismo en el codon 129 del gen PRP, este codon puede codificar metionina o valina y esto tiene influencia en el fenotipo de la enfermedasd inducida por mutaciones en otras posiciones (17). Además, se puede transmitir una encefalopatia espongiforme de tipo familiar (es decir de origen genético producida por mutación de PRP) a animales de laboratorio, por ejemplo por inyección intracerebral, en nervios periféricos e inoculación oral de preparaciones homogeneizadas de tejido del indiviuo que padece la TSE familiar, y esto demuestra que en estas enfermedades genéticas se forman particulas de prion infectivo prpsc.

Por otra parte, en las variantes esporádicas de estas enfermedades existe una conversión espontánea de prpc a prpsc, según se cree esto ocurre en ausencia de mutaciones en el gen y en ausencia de prpsc exogena (74). Algunas investigaciones sugieren que en realidad podria haber alguna mutación somatica no detectada que favorece la conversión de prpc a prpsc, en el tejido con dicha mutación, o que podria tratarse de mutaciones en algun gen que codifica para una proteína que interactua con prpc afectando su conformación, por ejemplo una chaperona. Otra posibilidad es el cambio espontaneo de prpc normales a la forma infecciosa en respuesta a algun cambio en el ambiente celular (por ejemplo alteraciones producidas por otras patologías) que afecte su conformación.

Principales lesiones que aparecen en el sistema nervioso

**Deposición de amiloides:**

En muchas encefalopatías espongiformes se observan en las neuronas mediante la coloración con rojo congo particulas con forma de baston similares a los amiloides. Al microscopio electrónico se observan como fibrillas conocidas como SAF (fibrillas asociadas a scrapie) y tambien como “prion rods”. Las partículas SAF se deben a la agregación de moléculas de Prpsc formando amiloides (85).

Los amiloides son agregados proteicos en forma de fibrillas de unos 100 A de diámetro compuestos por proteinas con conformación de laminas $\beta$ planas. En las TSEs estan compuestos por prpsc. Son insolubles y bastante resistentes a proteasas. Se tiñen intensamente con colorantes como el rojo congo y se acumulan en el exterior celular como consecuencia de la muerte celular por apoptosis.

Estos complejos son típicos de enfermedades neurodegenerativas tales como el Alzheimer y la enfermedad de Parkinson, aparecen tambien en las enfermedades causadas por priones. Ademas existen amiloidosis hereditarias, en las cuales se forman amiloides debido a la mutación en alguna proteína que entonces adquiere una conformación que favorece su agregacion con más moléculas de si misma (81).

Los amiloides no aparecen solo en el sistema nervioso de los mamíferos y otros vertebrados sino que serían bastante comunes entre los organismos eucariotas en general. Los priones de S.cerevisiae tambien los producen.

**Espongiosis**

Se manifiesta tanto en la sustancia gris como en la sustancia blanca del sistema nervioso central. Consiste en una intensa vacuolización del citoplasma de las neuronas, quedando el núcleo y otras organelas en la periferia celular . Es decir, se trata de la formación de vacuolas intracelulares y no extracelulares como originalmente se penso (estas aparecen en preparados de tejido nervioso pero constituyen un artefacto producido por el tratamiento de fijación) (66).

La vacuolizacion que se observa en las TSEs esta relacionada con la sobreeexpresión de acuoporinas 1 y 4 que producen el incremento del contenido de agua intraneuronal (86).

Otra lesión que puede observarse es la astrogliosis, que se puede detectar mediante la coloración de Nissl y Cajal. La astrogliosis es la hiperplasia e hipertrofia de los astrocitos en respuesta a daños en el sistema nervioso y uno de los marcadores bioquímicos de la misma es la presencia de alta concentración de la proteina glial acidica (GFAP) (87).

Tabla1. Enfermedades producidas por priones en humanos

| Enfermedad | Mecanismo de patogenesis |
|---|---|
| KURU | Puede aparecer por mutación somatica, tambien transmitirse en forma infecciosa (canibalismo) |
| Creutzfeld-Jacob esporadica (sCJD) | Aparece por mutación somatica al azar |
| CJD iatrogenica (iCJD) | Infeccioso, a traves de instrumentos quirúrgicos y jeringas infectados, tambien material biologico, por ejemplo tratamientos con hormona del crecimiento |
| CJD familiar (fCJD) | Genetica, hereditaria, es potencialmente transmisible en forma infecciosa. |
| CJD variante y nueva variante (vCJD, nvCJD) | Infeccion por medio de alimentos contaminados con prion BSE |
| Síndrome de Gerstman-Straussler-Scheinker (GSS) | Hereditaria, potencialmente infecciosa |
| Insomnio fatal familiar (FFI) | Hereditaria |
| FFI esporadico | Mutacion genetica al azar |
| Sindrome de Alpers | Mutacion genetica al azar |

Tabla 2. Enfermedades producidas por priones en animales.

| Enfermedad | Hospedador | Mecanismo de patogenesis |
|---|---|---|
| Scrapie | Ganado ovino y caprino | Puede ser genetica, esporadica por mutacion al azar y tambien infecciosa (ingesta de pasto contaminado, de las madres a las crias, etc) |
| Encefalopatia | Ganado bovino | Infeccioso, por ingesta de alimentos |
| espongiforme bovina (BSE) | | contaminados con el prion del scrapie |
| Encefalopatia espongiforme del vison | Visones de criadero | Infecciosa por ingesta de alimentos contaminados con scrapie y transmisible por mordeduras. |
| Encefalopatia de ungulados exoticos | antilopes | Infecciosa por alimentos contaminados con priones de la BSE. |
| Enfermedad del cansancio cronico (CWD) | Renos | Infeccioso |
| Encefalopatia espongiforme felina | gatos | Infeccioso por ingesta de alimentos contaminados con BSE. |

A continuación se describe brevemente cada una de estas enfermedades:

## Enfermedades causadas por priones en humanos

### Enfermedad de Creutzfeldt-Jakob

La enfermedad de Creutzfeldt-Jakob es una enfermedad neurodegenerativa fatal de incidencia muy baja en la población (aproximadamente 1 individuo en 1 millon). La forma tipica de esta enfermedad afecta generalmente individuos con edades de entre 50 y 75 años y se caracteriza por la presencia de espongiosis en la corteza cerebral y en el cerebelo, acompañado de gliosis marcada y deposición de amiloides formados por prpsc.

Recientemente se descubrio otra variante de la enfermedad que fue denominada nvCJD (new variant CJD) que se caracteriza por afectar a individuos mas jóvenes y se transmite por el consumo de carne contaminada con el prion de la encefalopatia espongiforme bovina. Esta variante fue la causante de una gran cantidad de casos en personas que consumieron carne de vacunos infectados en Europa en la decada del noventa y tuvo amplia repercusión en los medios. La vCJD presenta un cuadro clinico algo distinto al del CJD clásico, con vacuolización en los ganglios basales, marcada astrocitosis en el talamo, y amiloides rodeados por vacuolas (74).

La forma clásica de la enfermedad puede presentarse en tres formas distintas (que tambien estan presentes en todas las demas enfermedades producidas pro priones):

Esporádica : aparece por mutación somatica en el gen PRP o por cambio de conformación espontaneo de prpc.

Iatrogénica: esta forma de la enfermedad se contrae por contaminación con priones de material quirúrgico, endoscopios, electrodos cerebrales ,etc y tambien aparece por utilización de hormona de crecimiento extraida de cadáveres contaminada, transplantes de cornea, transplantes de dura madre entre otros.

Familiar: aparece por mutaciones en el gen PRP que se transmiten de generación en generación.

Infecciosa: esta forma, entre humanos no ocurre pues seria necesario el canibalismo como en el caso del kuru para que exista la transmisión. Se demostro su existencia mediante inoculación en animales de laboratorio (lo mismo se observó con las demas TSEs).

Entre los síntomas de todas las variantes de la CJD se incluyen una rápida y progresiva demencia, disfunción motora y alteraciones en el patrón electroencefalografico (88).

Esta enfermedad fue descripta por primera vez por H.G. Creutzfeldt en 1920 a partir de un paciente con síntomas atípicos con respecto a las enfermedades conocidas hasta esa epoca. En 1921, el neurólogo A. Jacob describio 4 casos, dos de ellos con características tipicas de la CJD (88). El nombre de esta encefalopatia deriva del de estos dos investigadores.

La forma familiar de la CJD se conoce desde hace bastantes años y se sabe que hay varias mutaciones en el gen PRP que codifica la proteina Prp capaces de producir CJD. Otras mutaciones en este gen se encuentran asociadas a la enfermedad GSS y al insomnio fatal. Las mutaciones de PRP se heredan en forma dominante (88).

<u>Enfermedad del insomnio fatal familiar (FFI)</u>

Se trata de una enfermedad genética autosómica dominante, causada por mutaciones en el gen PRP, precisamente debido a la sustitucion de un ac aspartico por asparagina en la posición 178 de la proteína (17,89).

La misma mutación esta asociada tambien a algunas formas de CJD familiar, sin embargo la diferencia entre ambas entidades radica en un polimorfismo metionina/valina en el codon 129. Los individuos heterocigotas presentan una tiempo de evolución mas prolongado que los homocigotas para metionina (19).

La enfermedad se caracteriza por una importante atrofia del tálamo. Las lesiones neuropatológicas de FFI incluyen astrogliosis, principalmente a nivel de las regiones ventral anterior y dorsal media del tálamo (18,89). Entre los síntomas se puede destacar la aparición de insomnio progresivo acompañado de perdida de las capacidades cognitivas (19).

<u>Síndrome de Gerstmann - Sträussler - Scheinker (GSS)</u>

La enfermedad de Gerstmann-Straussler-Scheinker es mucho menos frecuente que la CJD con una incidencia estimada en 2 casos por cien millones de habitantes. Los sintomas suelen aparecer entre la tercera y sexta década de vida en forma de un cuadro atáxico de curso progresivo al que se añade de forma tardía un deterioro cognoscitivo. La duración de la enfermedad oscila entre 1 y 10 años (19).

A nivel histologico se caracteriza por la presencia de espongiosis y de gran número de placas amiloides, muy notorias en el cerebelo (90,91,92).

El GSS fue transferido a primates y roedores por inoculación intracerebral, y a hamster por inserción del gen anormal PRP en el genoma del animal. (91,92)

Kuru:

Enfermedad confinada a varios valles adyacentes del interior montañoso de Nueva Guinea, donde afecta a unas 160 aldeas indígenas. Fue descrita por primera vez en 1957, por V. Zigas y C. Gajdusek. En aquella época, el kuru afectaba de forma epidémica a la tribu de los Fore, transmitiéndose por medio del canibalismo ritual que ésta tribu llevaba a cabo. Así, con la desaparición del canibalismo en Nueva Guinea, el Kuru ha experimentado un descenso progresivo (90,91,92,93).

Los principales síntomas del kuru son ataxia cerebelar, pérdida de coordinación, temblores (escalofrios que evolucionan hasta una completa incapacidad motora), congestión de los vasos sanguíneos, atrofia cortical y, posteriormente, demencia (90,91,92,93).

El periodo de incubación puede superar los 30 años, sin embargo, la muerte se suele producir en menos de 2 años a partir de la aparición de los primeros síntomas.

Síndrome de Alpers

Es una enfermedad crónica neurodegenerativa similar a la enfermedad de Creutzfeldt-Jakob pero con la diferencia de ademas de los tipicos signos neuronales, tambien produce degeneración en el higado. Solo afecta a niños y recien nacidos, con una bajisima frecuencia en la población. (90).

El síndrome de Alpers puede ser transmitido experimentalmente a hamsters por inoculación intracerebral, pero no a oras especies. (90).

## Enfermedades producidas por priones en animales:

### Enfermedad del Scrapie

Esta enfermedad se conoce desde el siglo XVIII. Actualmente se encuentra ampliamente extendida en Europa, Asia y América siendo la encefalopatia espongiforme mas comun. Es una enfermedad neurodegenerativa del sistema nervioso propia del ganado ovino.

En 1938 fue demostrado su carácter tranmisible en experimentos de J. Cuillé y P.L. Chelle quienes inocularon con scrapie de animales enfermos a cabras sanas, que luego desarrollaron la enfermedad.

En 1939, el veterinario norteamericano W.J. Hadlow señaló la similitud que existe entre el Scrapie y una enfermedad humana neurodegenerativa (el Kuru).

A principios de los años cincuenta se hizo patente la naturaleza atípica del supuesto patógeno por su extraordinaria resistencia a agentes fisicoquímicos que normalmente inactivan a los virus. En esta época se constató la diferente susceptibilidad al scrapie entre distintas razas de ovejas y en 1965 se estableció el concepto de la barrera interespecífica en la transmisión del scrapie. A partir de entonces se puso en duda la existencia de un ácido nucleico específico .proponiéndose la asociación del scrapie con una pequeña proteína básica y especulándose sobre la posible replicación de las proteínas. Finalmente, en 1982 Prusiner propuso la hipótesis del prion como una partícula infecciosa de naturaleza proteica responsable del scrapie (74).

Los signos clínicos más significativos del scrapie son temblores y un fuerte prurito que conduce a los animales a rascarse (de donde proviene el nombre de la enfermedad).

A nivel histologico la enfermedad presenta astrogliosis, intensa vacuolización intracelular y pérdida neuronal (94).

El prion causante del scrapie se puede transmitir a la descendencia del animal afectado a traves de la placenta asi como tambien a otros ovinos a través del contacto con pasturas contaminadas con el prion. Los síntomas de la enfermedad aparecen normalmente entre los 2 y 5 años de edad. Los animales infectados mueren alrededor de seis meses después de la aparicion de los sintomas clinicos.(74)

**Encefalopatía espongiforme bovina (BSE)**

Tambien conocida como enfermedad de las vacas locas, es una enfermedad bastante reciente. Surgio en el Reino Unido en al decada del 80 , desde donde se extendio a otros paises europeos. Se origino debido al uso de harinas fabricadas con restos de ovejas presumiblemente infectadas con scrapie (66).

Los animales infectados desarrollan transtornos en la locomoción y en el comportamiento tales como posturas anormales, agresividad, nerviosismo, dificultad para alimentarse e incorporarse (66).

En la BSE aparece vacuolización bilateral simétrica en el tronco encefálico(74).

**Encefalopatía Espongiforme Felina (FSE)**

Fue diagnosticada por primera vez en 1990 en un gato doméstico y a partir de entonces se encontraron numerosos casos en estos animales, principalmente en el Reino Unido. Se cree que esta encefalopatia espongiforme es una enfermedad nueva que se originó en los felinos a partir del prion de la enfermedad de las vacas locas contaminante de los alimentos balanceados para mascotas. Esto es coherente con el hecho de que los primeros casos aparecieron en plena expansion de la epidemia de BSE en el ganado vacuno y ademas, en estudios histologicos de los animales muertos por la enfermedad se observaron los signos tipicos de la encefalopatia causada por el prion de la BSE (66).

**Síndrome del cansancio crónico (CWD)**

El CWD (chronic wasting disease) afecta animales salvajes de la familia de los cervidos , especialmente alces y renos, pero tambien aparece en antílopes.

Los animales afectados permanecen postrados, con incordinacion de los movimientos, dejan de alimentarse y finalmente mueren (90).

**Encefalopatia transmisible del vison (TME)**

Es similar a la BSE y al CWD, se descubrio en criaderos de visones destinados a la industria peletera. Una hipótesis sobre su origen sugiere que aparecio en estos animales de una manera similar al de la epidemia de BSE, por el consumo de alimento contaminado con priones del scrapie o de la BSE (74,90).

## Diagnostico de las TSEs

El método más seguro para el diagnóstico de una enfermedad priónica consiste en la detección inequívoca de la proteína PrPSC por lo que las técnicas basadas en la utilización de anticuerpos especificos, (principalmente Western blot e inmunohistoquimica) , son las mas empleadas en el diagnostico. Actualmente se dispone de anticuerpos monoclonales que distinguen entre prpsc y prpc (94).

El diagnostico postmortem de TSEs se puede realizar por la técnica de Western blot con anticuerpos que reconocen prpsc (96). Para esto, se obtienen muestras de tejido cerebral, se homogenizan y luego se trata con proteinasa K, que degrada cualquier proteína que no sea prpsc, que es proteasa resistente. A continuación se realiza una extracción de proteínas y la fracción proteica se corre en un gel de poliacrilamida. Las bandas del gel se transfieren a una membrana de nitrocelulosa donde se revela la presencia de prpsc mediante anticuerpos específicos comerciales. En este ensayo se basa un test comercial muy utilizado para la detección de TSE en animales denominado test Prionics. Este test emplea muestras de tejido cerebral de animales enfermos, de animales clínicamente sanos y de otros con enfermedades neurológicas no relacionadas con los priones y la mayor desventaja que presenta es que su sensibilidad no es suficientemente alta como para detectar las isoformas patológicas en muestras de animales vivos, por lo que únicamente se pueden analizar muestras de cerebro tras el sacrificio.
Otros tests que se están empleando en la actualidad tambien se basan en la resistencia a proteasas que presentan los agregados de prpsc y en la detección con técnicas de Western blots de las formas resistentes (94,95).

También se puede detectar PrPcs en secciones de tejido embebido en parafina y fijado con formalina. Para revelar la presencia de prpsc se pueden emplear diversos pretratamientos. Uno de los más simples y efectivos para la tinción de la proteína, y tambien de agregados amiloides, es el tratamiento con 90% de ácido fórmico en secciones deparafinizadas .

Ademas, se puede utilizar microscopia electrónica para detectar fibrillas asociadas a scrapie (SAF) por microscopia electrónica que permiten diagnosticar el scrapie (96).

Los resultados obtenidos mediante las técnicas anteriores son validados mediante el bioensayo del homogenato cerebral que supuestamente posee priones en ratón. Esta técnica permite identificar la cepa de prion presente en el homogenato, pues estan estandarizadas (96). Existen diversas líneas de ratón que expresan distintas cepas de prpc y debido a la existencia de la barrera de especies y al reconocimiento entre cepas, los priones de la muestra solo produciran enfermedad en el ratón si pertenecen a la misma cepa que la prpc expresada en el sistema nervioso del animal.

También se pueden diagnosticar TSEs mediante el análisis de líquido cefalorraquídeo para determinar la presencia de proteínas marcadoras de estas enfermedades, esto es, proteinas cuya expresión esta alterada en estas enfermedades pero no en otras. Uno de los marcadores mas utilizados es la proteina 14-3-3 (94,95).

La familia de proteínas 14-3-3 está compuesta por 7 isoformas. Fueron descubiertas en 1967 tras un estudio intensivo de proteínas de cerebro bovino y recibieron esta nomenclatura, que todavía se mantiene, por sus perfiles de elusión cromatográfica y de movilidad electroforética. Son proteínas de distribución ubicuas, se encuentran en todos los tejidos de todos los organismos eucarióticos estudiados y están implicadas en la regulación del ciclo celular y de la apoptosis mediante su unión a motivos estructurales de proteínas reguladoras específicas. La presencia de la proteína 14-3-3 en líquido cefalorraquídeo permite discriminar TSEs de otras enfermedades neurológicas pues su expresión esta aumentada en ls encefalopatias espongiformes. Aun asi, hay casos en los que el diagnostico de una TSE mediante este marcador no es definitivo pues al tratarse de una proteína que se expresa en muchos tipos celulares en el sistema nervioso, puede ser liberada al liquido cefalorraquideo en enfermedades infecciosas que producen muerte celular, por ejemplo meningitis bacterianas (94,95).

## Aspectos terapeuticos:

Actualmente no existe ningún tratamiento para las TSEs y se están realizando numerosas investigaciones enfocadas al diseño de fármacos para evitar la formación de la isoforma prpsc o para facilitar su eliminación. Probablemente, los avances en el estudio de enfermedades como el Alzheimer favorezca el desarrollo de fármacos efectivos frente a los priones. No hay que olvidar que ambos tipos de enfermedad afectan al sistema nervioso y producen agregados amiloides en las neuronas.

Entre los compuestos ensayados para el diseño de fármacos contra los priones se encuentra el colorante rojo congo, un pentosan sulfato. Se sabe que bloquea la formación de amiloides, pero es toxico para las celulas por lo que habría que encontrar algun otro pentosan sulfato que no lo sea. De todos modos, no es capaz de bloquear la accion de prpsc una vez que la enfermedad esta avanzada. Aparentemente solo es eficaz si se lo administra a ratones al mismo tiempo que se los infecta con el prion (97).

Una alternativa son las antraciclinas como la doxorrubicina y sus derivados. Para demostrar su eficacia se realizaron ensayos en hamsters a los que se les administraba simultáneamente el fármaco y homogenatos de cerebro de ratones infectados con PrPSC. En estos estudios se encontro que se retrasa la aparición de los síntomas clínicos de la enfermedad y se prolonga significativamente el periodo de supervivencia de los ratones. Además, el derivado 4-iodo-4desoxi- doxorrubicina bloquea la formación de amiloides in vitro (97).

Algunos tetrapirroles ciclicos inhiben la formación de prpsc in vitro y al ser ensayados in vivo en ratones infectados con prpsc la supervivencia se incrementa entre el 50% y el 300% . Estos compuestos estabilizan la conformación normal de prpc (impiden su cambio conformacional en presencia de ppsc) por lo que serían buenos candidatos para el diseño de terapias contra las TSEs (99).

Otro enfoque experimental consiste en disminuir los efectos negativos de la depleción de prpc producida por accion de prpsc que le cambia su conformación volviéndola no funcional. Uno de los fármacos que podria ser util para esto es la flupirtina, un analgésico no opioide (94,95). In vitro tiene un fuerte efecto protector de neuronas infectadas con PrPSC porque incrementa los niveles intracelulares de la proteína antiapoptótica Bcl-1 y del antioxidante glutation (94,95).

Otros compuestos analizados son los Glicanos polianiónicos y sulfato de dextrano, que estimulan la endocitosis de prpc evitando que tenga acceso a prpsc (74).

Probablemente el mejor enfoque terapéutico consista en el bloqueo de la interacción entre prpc y prpsc mediante el empleo de algun fármaco que interfiera con dicha interacción, por ejemplo se podrían diseñar alguno fármaco que mimetice a prpc de manera tal que prpsc tenga mas afinidad por el fármaco que por la verdadera prpc.
Con respecto a esto, en un estudio con ratones que expresan prpc fusionada a un fragmento Fc de inmunoglobulina (prpc soluble dimerica) se vio que la proteína de fusión inhibía la replicación de prpsc evitándose asi la enfermedad. Dicha proteína funcionaba como un receptor de prpsc que era capaz de unirse a esta pero resultaba resistente al cambio conformacional con lo cual prpsc no podia propagarse (100).

## Mecanismos de patogenesis de prpsc

La patogénesis de los priones es un proceso complejo que puede dividirse en las siguientes 4 fases que son: infección, replicación en los tejidos periféricos, transmigracion desde los tejidos periféricos al sistema nervioso central y neurodegeneración.

Infección y replicación en los tejidos periféricos:

Entrada por via oral y transporte al sistema nervioso central:

El evento clave en la patogénesis de los priones es la conversión de prpc en copias de la isoforma patógena prpsc que como ya indiqué anteriormente, es proteasa resistente.
En rasgos generales, la entrada de prpsc por via oral es capaz de iniciar la reacción de cambio conformacional, convirtiendo las moléculas de prpc intestinales en mas prpsc, que invade entonces los nervios periféricos cercanos, que también expresan prpc, con lo cual se va propagando hasta que llega al sistema nervioso central. Este transporte es muy lento y depende de la cantidad inicial de prpsc que llega a las terminaciones nerviosas y de la expresión de prpc. Ademas no es el unico mecanismo existente: tambien se ha propuesto transporte a traves de los vasos sanguíneos mediante unión a proteinas plasmáticas e infección de celulas por contacto directo. El hecho de que prpsc es resistente a proteasas favorece la infección por via gastrointestinal, pues queda protegida de la degradación por enzimas digestivas.

En modelos murinos de TSEs se detecta la presencia de prpsc en el sistema nervioso entérico (SNE) luego de la inoculación oral de los priones, esto constituye evidencia a favor de un mecanismo de entrada del patógeno por via oral seguida de transporte a las terminaciones nerviosas del intestino (101). Diversos estudios indican acerca de posibles receptores para prpsc en la mucosa intestinal y aparentemente el principal seria el precursor del receptor de laminina de 37 kD (102), este receptor se incorpora al receptor maduro que tiene 37kD y a partir de estudios de inmunohistoquimica con anticuerpos monoclonales en secciones de tejido en parafina se descubrio que se expresa en el ribete en cepillo del intestino delgado en humanos (103).

A partir del empleo de anticuerpos contra prpc en secciones de biopsias de tejido intestinal se demostro que prpc colocaliza con los filamentos proteicos neurales de las fibras nerviosas entericas (104). Experimentos de hibridación in situ confirmaron la presencia del mRNA de prpc en celulas gliales de los ganglios entericos. Las terminaciones nerviosas del sistema nervioso enterico que expresan prpc se encuentran asociadas íntimamente a las celulas epiteliales del intestino. El receptor de laminina de 37 Kda se expresa en dichas celulas, ademas solo una capa epitelial delgada separa a las moléculas de prpsc en el lumen intestinal de las terminaciones nerviosas (102).

Un modelo que explica como llega el prion al sistema nervioso postula que luego de la ingestión, prpsc llega al epitelio intestinal donde entra en las células a traves de endocitosis mediada por el precursor del receptor de laminina de 37 kD. Una vez internalizada, prpsc convierte prpc en mas prpsc, y es transportada al sistema nervioso entérico, esto involucra transportes celula-celula.

Las moléculas de prpsc en el intestino ingresarian en primer lugar a las células M intestinales que son células del epitelio de las placas de Peyer cuya función es la presentación de antígenos a los linfocitos y otras células inmunes. Los antigenos los obtienen del lumen del intestino y los incorporan mediante el mecanismo de transcitosis.

En experimentos con lineas celulares de células M creciendo en monocapa junto a una línea celular de epitelio intestinal (caco-2), se observa transporte activo depeniente de la temperatura de bolitas de latex conjugadas a FITC con priones infectivos (105). Como resultado se encontro que el prion es transportado a traves de la monocapa celular. Por otra parte, no se observa transporte de priones en cultivos de celulas caco-2 en ausencia de celulas M.

La transcitosis dependiente de celulas M permite a los priones acceder a celulas inmunologicas, en especial celulas dendríticas. Estas forman una capa celular densa en las placas de peyer en regiones que estan en intimo contacto con el epitelio asociado que contiene las celulas M. De esta manera, las células dendríticas pueden infectarse con prpsc proveniente de las células M. De hecho, células dendríticas in vitro pueden adquirir prpsc y como existen subpoblaciones de estas células que son migratorias, pueden propagar el prion adquirido a traves del tejido linfático (106).

Ademas, el uptake de prpsc en el lumen del intestino esta restringido a estas células y no se encuentra prpsc en otras células inmunologicas.

Las celulas dendriticas foliculares expresan prpc y esto las hace criticas para la replicación del prion en el tejido linfoide y posterior neuroinvasion ya que se necesita prpc para poder obtener mas copias de prpsc. La pérdida temporal de la diferenciacion de estas celulas mediante expresión de una proteina de fusion que consiste en una inmunoglobulina fusionada al receptor B de linfotoxina (LTB R- ig) antes de la inoculación intraperitoneal de scrapie en ratones bloquea la acumulación temprana de infectividad en el bazo y reduce la susceptibilidad a la enfermedad. Además, el tratamiento con LTBR-Ig luego de la inoculación oral de scrapie bloquea la acumulación de prpsc en las placas de peyer y nódulos linfáticos mesentericos, evitando la neuroinvasion (82,107).

## Transporte al sistema nervioso y mecanismos de neurodegeneracion:

Prpsc llega al sistema nervioso central a traves de la migración desde organos perifericos en especial el bazo, intestino y sistema linfoide y tambien a traves de su propagación por los nervios periféricos (107), en forma similar a la propagación de virus herpes, es decir, avance a traves de la infeccion de terminaciones nerviosas y avance a traves del nervio a medida que se va replicando utilizando para esto las moléculas de prpc que encuentra a su paso. Este proceso es muy lento y tendría relación con el periodo de incubacion de las TSEs (66). La propagación de prpsc por los tejidos periféricos también ocurre a traves de la infección sucesiva de las células que expresan prpc que encuentra a su paso (células nerviosas, linfocitos B, células dendríticas foliculares, keratinocitos, etc.) el prion es capaz de pasar de célula a célula. No se sabe bien como ocurre esto. Probablemente el mecanismo implicado sea el de endocitosis mediada por receptor de moléculas de prpsc liberadas al espacio intercelular, otra posibilidad es que prpsc aproveche algunos de los transportadores existentes como por ejemplo los Gap junctions que comunican directamente celulas entre si. La expresión de prpc en la celula blanco de prpsc es absolutamente necesaria para que exista infección. Un sitio de replicacion y acumulacion de prpsc es el bazo y tejido linfoide asociado (82). Las células dendríticas son muy importantes para la replicación de prpsc. En cambio, los linfocitos T no estan involucrados. La timectomia no altera el tiempo de incubación de una TSE y ratones deficientes en celulas T no son menos suceptibles a las TSEs ni a la acumulacion de prpsc en tejidos perifericos (82). La deficiencia de celulas B interfiere con la replicación de prpsc en el bazo y con la neuroinvación. Evidentemente dichas celulas transportan prpsc largas distancias y de esta manera es más probable que llegue más rápido al sistema nervioso, acortandose el tiempo de incubación, que de por si es prolongado aún en presencia de esas células pero mucho más en su ausencia. De todos modos, las células dendríticas foliculares son las más importantes para la replicación de prpsc de todas las celulas inmunes (82). En cuanto a los macrófagos, contribuyen a inhibir la replicacion de prpsc en los tejidos endocitando células infectadas en los tejidos perifericos, en especial linfocitos Se puede detectar prpsc dentro de los lisosomas de los macrófagos, donde queda secuestrada y asi se retarda el tiempo de incubacion (82).

Por otra parte, es necesario tener en cuenta que la entrada del rprion no siempre ocurre por via oral, también puede transmitirse por cualquier otra vía que involucre un daño al organismo combinado con exposición al prion, por ejemplo se conoce la transmisión denominada iatrogénica, a través de instrumentos quirurjicos, electrodos cerebrales y también por medio de transplantes de tejido infectado principalmente cornea y dura madre (68).

Neurodegeneración:

Muchas enfermedades neurodegenerativas involucran el procesamiento anormal de proteínas neuronales y la acumulación de proteínas mal plegadas, las TSEs no son la excepcion (10,81). La neurodegeneración producida por los priones es un proceso multifactorial que involucra acumulación de proteinas anormales en amiloides y agresomas, estrés oxidativo y apoptosis.

Para que prpsc reconozca a prpc ambas deben encontrarse en la misma membrana plasmática. Prpsc se une a prpc y se produce la reacción de cambio conformacional que propaga la infección pues aumenta la cantidad de prpsc. En esta reacción puede estar involucrada alguna chaperona celular u otro tipo de molécula como se explicó anteriormente.

La reaccion puede ocurrir en cualquier superficie de membrana que contenga prpc y prpsc o a la cual prpsc tenga acceso (por ejemplo, prpsc en una vesícula endocitica que sale del RE puede convertir a la prpc presente en el golgi cuando dicha vesícula llegue a esta organela).

Una vez que prpsc llega a una neurona y accede a prpc, le cambia la conformación. Dado que prpc se ubica en dominios de membrana donde se produce endocitosis, el complejo resultante de prpsc puede ser endocitado y ser dirigido a los lisosomas. Una vez alli, es atacado por las proteasas lisosomales y como prpsc es parcialmente resistente no podra ser totalmente degradado. La acumulación de esta proteína parcialmente degradada en los lisosomas puede desencadenar el aumento del daño oxidativo debido al esfuerzo celular para degradarlas y además puede desencadenar una señal de apoptosis. Por otra parte, el estrés oxidativo puede estar aumentado también por la depleción de prpc producida por su conversión a prpsc con lo cual ya no puede cumplir su funcion antioxidante (33) (prpc esta involucrada en la defensa frente a radicales libres dada su actividad de superoxido dismutasa o su relacion con estas a traves de su funcion en el metabolismo del cobre).
Se sabe que en las TSEs existe un aumento pronunciado de la concentración de oxidantes como los radicales libres en el cerebro(33).

Por otra parte, es muy probable que durante la degradación parcial de prpsc en los lisosomas se forme el fragmento prp 27-30. Dicho fragmento es neurotóxico y desencadena apoptosis. Esto fue observado en varios estudios. Ademas, se demostró que dicho fragmento no solo es capaz de formar agregados amiloides citotóxicos sino que puede acomplejarse formando canales en membranas y estos canales permiten el paso de iones calcio (108). Este resultado podría tener gran importancia en los mecanismos por los cuales el prion produce daño celular. Los productos de degradación lisosomal de prpsc como por ejemplo el prp27-30 podrian escapar del lisosoma mediante algun mecanismo desconocido y acceder al retículo endoplásmico donde son capaces de agregarse formando canales permedables a iones o bien podrían acumularse en el citoplasma. En principio, fragmentos de prpsc como el mencionado tienen ancla de GPI, por lo que no podrían acumularse en el citoplasma sino que quedarian anclados a la membrana lisosomal luego de su degradación. Pero entre las enzimas lisosomales podrían encontrarse fosfolipasas capaces de cortar el GPI, con lo que se obtendría prp27-30 no anclada en solución capaz de llegar al citoplasma por algun mecanismo.

La formación de canales en la membrana del retículo por parte de prpsc se observó en células en cultivo. En otros estudios se encontró que la apoptosis inducida por priones ocurre a traves de la liberación de calcio del retículo endoplasmico y activacion de la caspasa 12 residente en dicha organela (109). El aumento de la concentración de calcio en el citosol es una señal de apoptosis, ademas, las caspasas son las proteasas efectoras de este mecanismo, degradando proteinas celulares. prpsc podría acceder a las moléculas de prpc que se estan sintetizando en ribosomas asociados al retículo mediante un mecanismo desconocido. Una vez en el retículo, prpsc interfiere con el correcto plegado de prpc supervisado por chaperonas residentes en el retículo y produce agregación proteica (109). Se sabe que en la patogenia de los priones algunas chaperonas favorecen la conversión de prpc en prpsc cuando esta ultima esta presente (este es el caso de la proteina X hipotética propuesta en los primeros trabajos sobre priones que se cree que es una chaperona), mientras que otras protegen a prpc de la accion de prpsc. En el retículo se encuentra la chaperona Bip (también denominada Gpr78) que tiene una funcion protectora para la célula infectada con prpsc. Bip se une a proteínas mal plegadas o agregadas y las dirige hacia su degradación en proteasomas. Se cree que en las TSEs familiares (producidas por mutaciones en el gen PRP que resultan en prpc mutantes que se pliegan mal en el retículo o se agregan) el largo periodo de incubación hasta la aparicion de los síntomas se debe a la acción protectora de Bip que media la degradación de las moléculas de prpc mutadas a traves del proteasoma (38). Como consecuencia de la agregacion de proteínas en el retículo, llega un punto en que la defensa dada por chaperonas como Bip no da abasto y entonces se desencadena la apoptosis a traves del mecanismo de respuesta a estrés en el reticulo, que se denomina respuesta UPR (unfolded protein response). El estrés en el reticulo es desencadenado por factores tales como la deprivación de nutrientes, o energia, flujos de calcio y acumulación de proteinas mal plegadas (110). Estos factores inician la respuesta UPR que se basa en la unión de Bip a las proteinas mal plegadas en forma dependiente de ATP.
Bip se une tanto a proteínas mal plegadas como también a las que estan siendo sintetizadas, pero la unión es mas persistente con las proteínas mal plegadas. La unión a las proteínas normales es transiente. En condiciones de ausencia de estrés, Bip se une a tres proteínas sensor transmembrana del retículo endoplasmico con lo cual produce la regulación negativa del UPR porque mantiene a dichos sensores inactivados. En presencia de estrés en el retículo (presencia de proteínas mal plegadas y/o agregadas), Bip se disocia de los sensores y se une a las proteínas mal plegadas, entonces, ls proteínas sensor liberadas permiten que el UPR se produzca, disminuye la síntesis general de proteínas en ribosomas asociados al retículo y se aumenta la biosintesis de chaperonas del reticulo. A continuacion se activa una vía de degradacion de las proteinas mal plegadas denominada ERAD (ER asociated degradation) que las dirige al proteasoma.
Bip monitorea la calidad de las proteínas en el retículo y regula cuales seran degradadas y cuales no. Además, el aumento en la síntesis de chaperonas del reticulo es acompañada por la liberación de calcio desde el reticulo al citosol, esto funciona como una señal apoptotica probablemente a través de la vía del inositol trifosfato (IP3) clásica (78,109) e involucra la activación de la procaspasa 12 residente en el retículo (esta es clivada originando la caspasa 12 activa)
La caspasa 12 va al citoplasma donde activa a la caspasa 3 que es central en los mecanismos de apoptosis, tambien se forman unas estructuras en el citoplasma denominadas agresomas que son complejos proteicos compuestos por agregados de prpsc, chaperonas Hsp70, subunidades del proteasoma, ubiquitina y vimentina. Se trata de complejos que se acumulan como resultado del intento de las celulas por degradar a prpsc. La formación de los agresomas activa las caspasas 8 y tambien la 3, esto conduce a la apoptosis (111).

Por otra parte, no siempre, es prpsc quien produce la neurodegeneración. Una de las isoformas topologicas de prpc (CTMprpc ) es neurotóxica por si misma y existen casos de TSE en los cuales no se detecta prpsc pero si acumulación de ctmprpc que ejerce efectos neurodegenerativos (69). Algunas mutaciones en el gen PRP favorecen la estabilización en el retículo de la forma topológica transmembrana CTMprpc, esta no se acumula en el retículo sino que es transportada en vesiculas al Golgi, donde se acumula y conduce a las celulas a la apoptosis a traves de un mecanismo desconocido (69).

Prpsc no solo se acumula en reticulo donde desencadena la respuesta UPR, sino tambien en lisosomas y citoplasma, al igual que el fragmento resultante de la degradacion con proteasas Prp27-30. El mecanismo de toxicidad que involucra lisosomas es desconocido, pero se sabe que conduce a la apoptosis a traves del efecto tóxico de la acumulación de fragmentos de prpsc parcialmente degradados en los lisosomas. En el citoplasma, la toxicidad de prpsc o de sus derivados de degradación involucraría la disfunción de la vía de degradación de proteínas citosólicas a través de proteasomas que reconocen proteínas marcadas con ubiquitina (111), además, algunos autores proponen que en algunos casos, en ausencia de prpsc y de ctmprpc puede aparecer neurodegeneración y TSEs, pues algunas mutaciones de prpc que rpesentan alteraciones queafectan su transporte a través de la vía secretoria pueden acumularse en el citoplasma y agregarse en amiloides, inhibiendo la degradación por medio de proteasomas (111).

## Priones de levadura y otros hongos

Una manera en que un prion de levadura (o cualquier otro ) puede manifestarse es a traves del cambio en una proteína codificada en un gen cromosómico, de forma tal que la forma alterada de la proteína es necesaria para la generación de mas moléculas de la misma forma a partir de la forma no alterada (es lo que ocurre con prpsc y prpc). Dicho cambio conformacional va a tener alguna consecuencia para la célula: si la forma infeciosa es tóxica (caso de prpsc) entonces la célula sera depletada de la forma normal y sufrirá apoptosis. Esto es lo que ocurre con el prion prp como explique antes. Ahora, si la conformación infecciosa es la forma inactiva de la proteína normal entonces en las células en presencia del priones infecciosos, la proteína normal se inactivará al convertirse en mas moléculas de la forma infecciosa y el fenotipo se sera el mismo que se obtiene en un mutante nulo para el gen de la proteína en cuestión. Este es el caso del prion Psi, que es la forma inactiva del terminador de la traducción sup35 y del prion URE3, que es la forma inactiva de ure2p (un regulador del catabolismo del nitrogeno).
Si la conformación infecciosa es la forma activa en alguna función celular, entonces el fenotipo dado por esta conformación es el wild type . Esto es lo que ocurre con el prion Het-s de podospora y con el prion beta de levadura. Ambos priones realizan funciones celulares normales y por esto la mayoria de las cepas de hongo wild type los tienen. En cambio, psi esta ausente en cepas wild type, su presencia seria desventajosa debido a la inhibicion de la traduccion (66). Aun asi se propaga eficientemente en la población pues siempre que se produzca un mating (cruzamiento) entre una levadura psi positiva con una negativa, toda la progenie sera psi positiva. La infección ocurre a traves del intercambio citoplasmatico entre las levaduras. Se trata de una herencia citoplasmatica con genes no cromosomales que además son proteínicos en forma análoga a la herencia de mitocondrias. En los hongos filamentosos, el intercambio involucra la fusión de hifas de colonias distintas y a traves de éstas pueden pasar los priones.

Los priones de levadura poseen tres caracteristicas que los distinguen de los virus:

"Curabilidad reversible": esto significa que la conformación infecciosa se puede eliminar por completo de las celulas infectadas mediante algun tratamiento y aun asi luego puede volver a aparecer en forma espontanea sin necesidad de fuente externa pues la proteína celular puede cambiar espontaneamente de conformación con cierta frecuencia baja originando de novo la conformación infecciosa. Esto con un virus no puede ocurrir, una vez que es eliminado de una celula no puede volver a aparecer en ella si no hay una nueva infeccion.

La presencia de la conformacion no replicativa es necesaria para la propagación de la conformación infecciosa .

La sobreexpresion de la forma no infecciosa incrementa la frecuencia con que esta aparece (es porque aumenta la probabilidad del cambio espontaneo.)

El prion URE3

Las células de levadura son capaces de reprimir la sintesis de las enzimas y transportadores necesarios para el crecimiento en medios pobres en ciertos nutrientes. Cuando crecen en medios ricos en nitrogeno como por ejemplo con glutamato, amonio, etc, la proteina Ure2p se une a los factores de transcripción Gln3p y Gat1p en el citosol impidiendo su transporte al núcleo con lo cual no pueden actuar sobre los genes involucrados en el metabolismo de compuestos pobres en nitrogeno como por ejemplo el alantoato y el ureidosuccionato (USA). Entre los genes que intervienen se encuentra el DAL5 que codifica para un tranportador de ambos compuestos mencionados y es inhibido por los factores mencionados (5).
Mutaciones en el gen ure2 que codifica la proteina ure2p no presentan la represion del los genes involucrados en el metabolismo de compuestos pobres en nitrogeno aun en presencia de compuestos ricos en nitrogeno como el amonio (1,5,112,113).
La proteína codificada por el gen ure2 puede existir en dos conformaciones, una de ellas es la proteína ure2p mencionada y la otra se denomina URE3 que tiene comportamiento infeccioso en forma análoga al caso de prp. En presencia de URE3, las levaduras presentan el mismo fenotipo que las levaduras mutantes nulas para el gen ure2 pues URE3 convierte a ure2p en mas copias de si misma.
URE3 posee las tres propiedades esperadas para los priones de levadura: puede se r eliminado de las celulas infectadas mediante tratamiento con guanidina pero de los clones curados se pueden aislar subclones que son URE3 positivos , La sobreexpresion de ure2p aumenta la producción de URE3, no por efecto del mRNA de ure2p sino por la presencia de la proteína (5,6) y por último, como indiqué, el fenotipo de las células URE3 positivas es el mismo que el de los mutantes nulos para Ure2p. Esto ocurre porque al propagarse URE3 mediante conversión de moléculas de ure2p, estas cambian de conformación tomando la conformacion infecciosa de URE3 y entonces ya no pueden cumplir su funcion (113).

Las células que presentan URE3 tienen también ure2p parcialmente resistente a proteasas y esta resistencia es característica de los amiloides (6) Ure2p se agrega específicamente en presencia de URE3 en forma dependiente de la secuencia de su extremo N-terminal (5,6). De hecho, el extremo N-terminal de 65 aa de Ure2p se denomina extremo prionico y es necesario para la agregacion y tambien para la innactivación de la proteína. Por otra parte, el dominio C-terminal de la proteína es el que cumple la función represora de la transcripcion. Ademas, el dominio prionico es capaz de agregarse en ausencia del dominio C-terminal y propagar a URE3. De esta manera, URE3 se propaga mediante la conversión de moléculas de ure2p en más moléculas de si misma que luego se agregan en amiloides.

## El prion PSI

Este prion tiene dos conformaciones: sup35 que es la version normal con función en la traducción y psi que es la forma infecciosa que se propaga a si misma mediante conversión de las moléculas de sup35. También forma amiloides como URE3 y su propagación puede ser tanto inhibida como favorecida por chaperonas como por ejemplo hsp104,hsp40 y hsp70 (114,115).

Sup35 es una subunidad de un factor de terminación de la traducción con funcion de reconocimiento de codones stop y clivaje del péptido del tRNA. En presencia de tRNAs supresores se establece una competencia entre el reconocimiento del codon de terminación por dicho tRNA lo cual permite insertar mas aminoácidos y el reconocimiento del codon de terminación por parte del factor de terminación, que resulta en la terminación de la síntesis. La presencia de la forma infecciosa psi depleta a las celulas de sup35 y de esta manera la síntesis de proteínas continua más alla del codón de terminación. (116,117). Al igual de lo que ocurre en el caso de ure2p, el extremo C-terminal de sup35 es el que se encarga de al función celular normal, es esencial para el crecimiento celular y presenta homologia con el factor de elongación de la traduccion EF-1 (112) mientras que el dominio N-terminal es el que reconoce psi y esta involucrado en el cambio conformacional y capacidad de formar amiloides. Este dominio, in vitro es capaz de agregarse en presencia de psi y se lo denomina dominio prion, es tipico de varios priones de levadura. Mutaciones en este dominio pueden impedir la propagación de psi (112) por otra parte el dominio C-terminal es funcional por si solo en sistemas de traducción in vitro (116,117).

El dominio prionico N-terminal de sup35 al igual que el de ure2p es rico residuos de glutamina y asparagina. Además, los residuos de glutamina son importantes para la propagación de la conformación infecciosa (118) y el dominio posee 5 repeticiones de octapeptidos similar al del prion prp (6). La deleción de repeticiones inhibe la propagación de psi, mientras que la adición de nuevas repeticiones en el dominio aumenta la frecuencia de aparición de la conformación psi de novo (119).

En la figura 3 se muestra un esquema resumido de los mecanismos de accion de los priones PSI (figura 3a) y URE3 (figura3b y c)

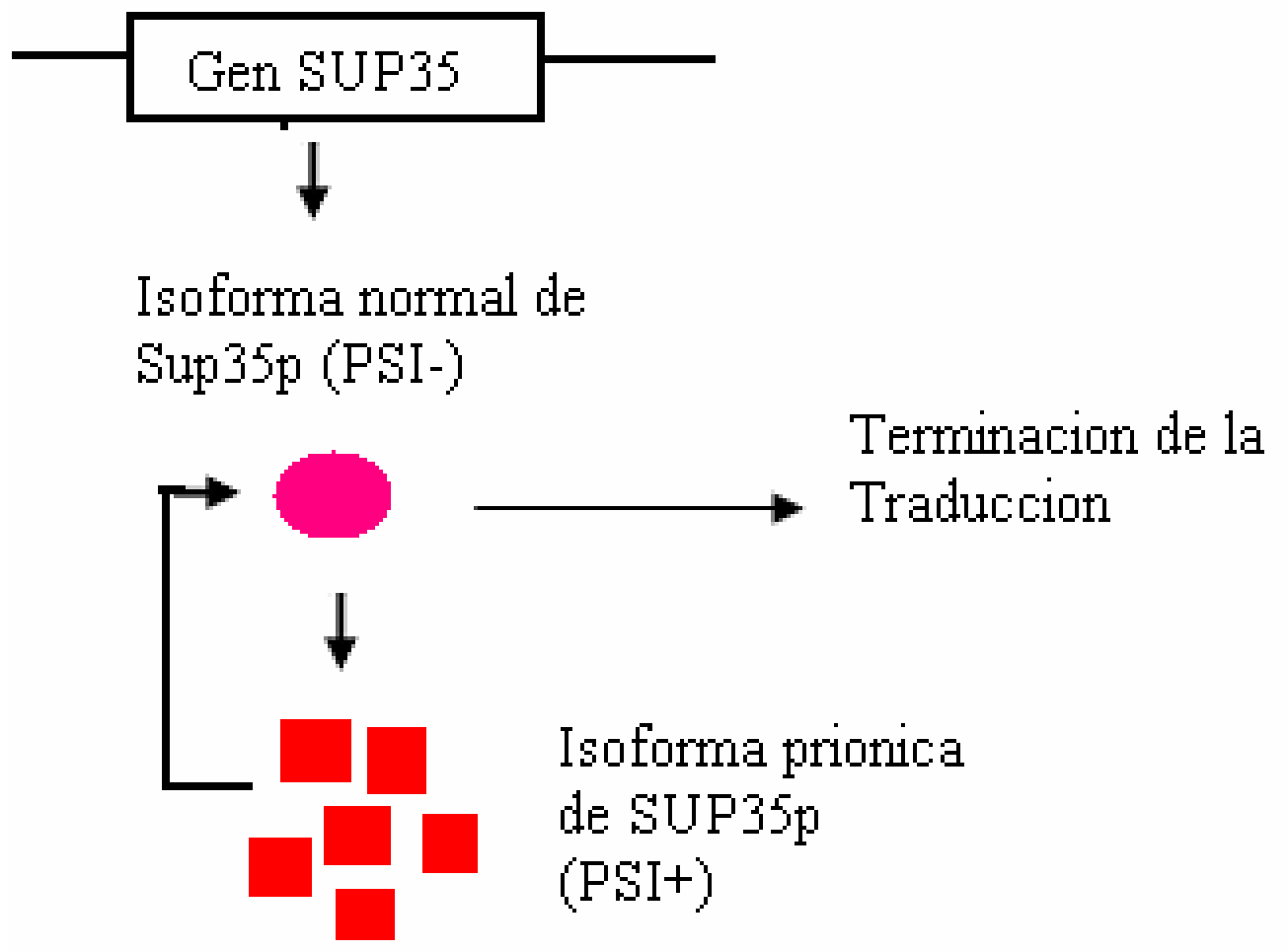

Figura 3. Algunos priones de levaduras y sus mecanismos de accion.
a) Psi secuestra moleculas de sup35 (isoforma normal), que pasan a formar parte del amiloide en crecimiento (cuadrados rojos). De esta forma sup35 no puede cumplir su funcion en la terminacion de la traduccion

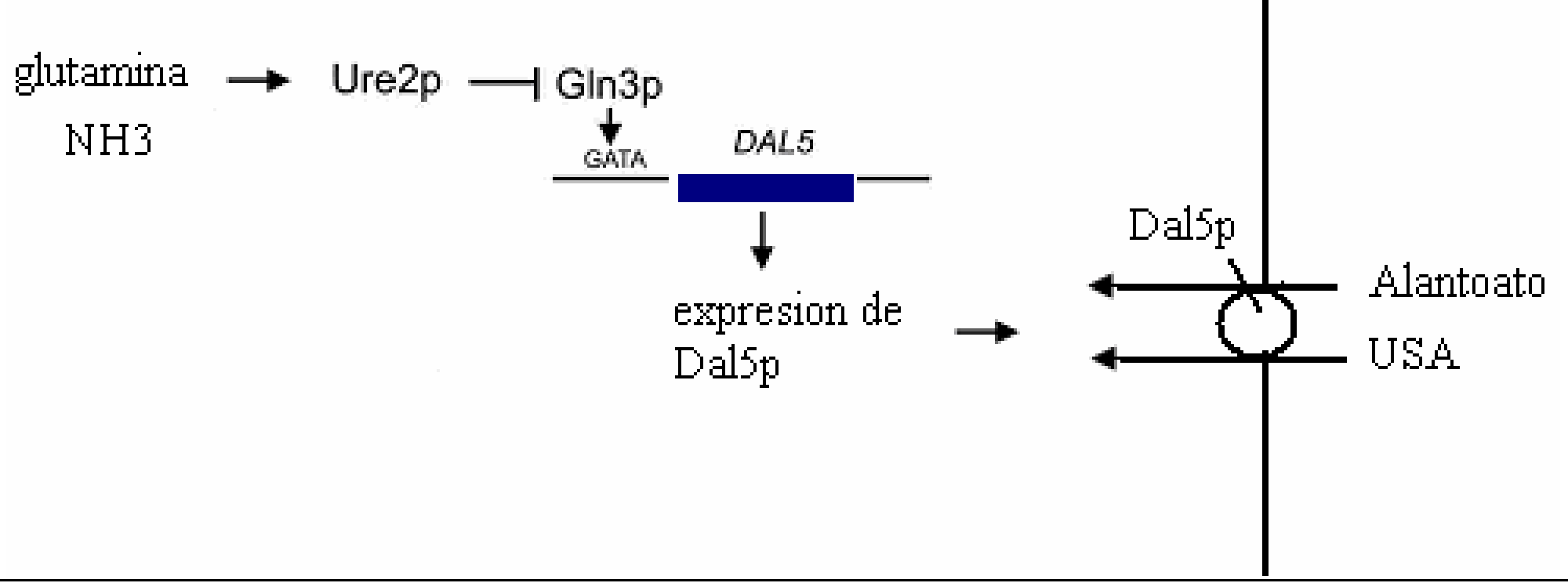

b) Ure2p regula el metabolismo del nitrogeno. Cuando las celulas crecen en medio con nitrogeno, ure2p se une a gln3p, impidiendo la transcripción de genes necesarios para el crecimiento en medios pobres en nitrogeno, como por ej Dal5p, que es un transportador de alantoato y de ureidosuccinato (USA). El alantoato es una fuente pobre de nitrogeno, mientras que USA es un intermediario de la síntesis de uracilo.

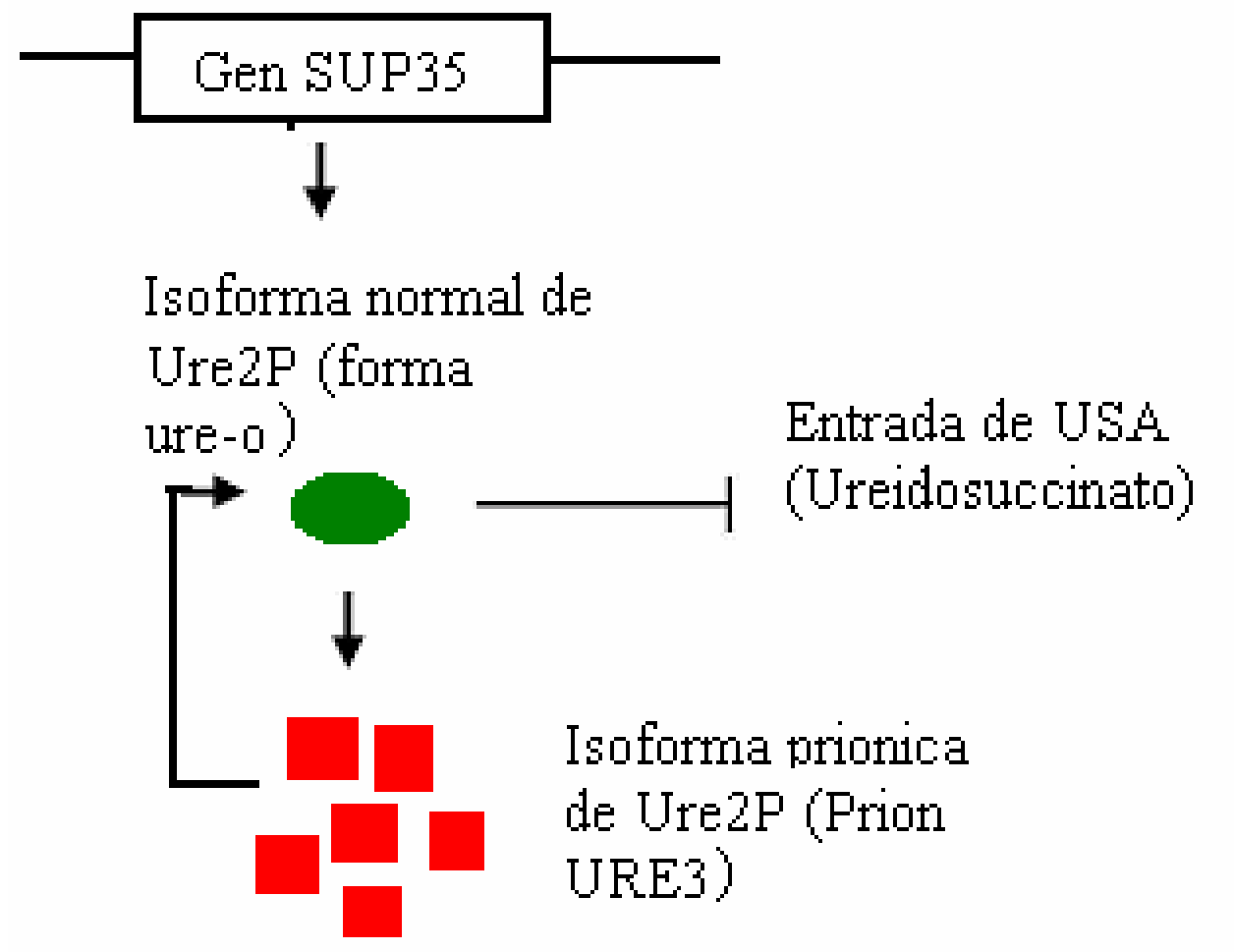

c) El prion URE3 secuestra moleculas de ure2p en amiloides. De esta manera disminuye la cantidad de moleculas disponibles para la regulación del metabolismo del nitrogeno (6).

## Otros priones:

## PIN

Es un prion que determina la inducibilidad del prion psi. La habilidad de la sobrexpresion de sup35 de inducir la aparición espontanea de psi es mayor en presencia de PIN. Existe en dos formas, la conformacion normal se denomina rnq1p y esta codificada en el gen RNQ1 y la conformación prionica (pues es la que se puede propagar) es PIN. En presencia de PIN, la proteina rnq1p se agrega en amiloides y toma la conformacion de PIN (6).

### Het-s

Interviene en la incompatibilidad citoplasmatica en podospora, es decir, cumple una función normal. Cada cepa de dicha especie posee su propia cepa de Het-s, que se diferencia de la de las demas cepas en solo 13 de sus 289 aminoacidos.
Las cepas de Het-s se comportan como alelos no cromosomales. El mating solo puede ocurrir entre colonias de podospora con el mismo tipo de het-s (5,6)

### Beta.

Se trata de un prion con función enzimatica. La forma infecciosa se denomina beta y es la forma activa de la proteasa vauolar beta. Esta conformación es esencial para activar al precursor inactivo. La presencia de la forma beta es esencial para las celuas de levadura, en su ausencia son incapaces de esporular y mueren mas facilmente al ser cambiadas de medio (6,119).

## Conclusion:

Los priones son proteínas únicas, pues son capaces de ingresar en sus hospedadores a traves de sus “portales naturales” como por ejemplo el sistema digestivo y asi infectar como lo hacen virus y bacterias, ademas son portadores de información genética en forma de conformaciones de proteína en lugar de ácidos nucleicos y esta variabilidad conformacional podria intervenir en la adaptación de las especies hospedadoras, no solo en la producción de enfermedad.

Las proteínas involucradas en enfermedades neurodegenerativas como el Alzheimer, Parkinson, Huntington, etc, forman amiloides similares a los que forman los priones, pero no se comportan como priones y hasta ahora no fue posible determinar si lo son (120). Estas proteínas producen lo que se denomina enfermedades conformacionales, pues son producidas por proteínas con conformacion alterada.
No esta demostrado que las proteínas responsables de dichas enfermedades puedan autopropagarse como los priones, pero en algunos casos parece que pueden ser transmitidos pues en experimentos en los que se inoculan amiloides en animales que ya presentan una amiloidosis la enfermedad se acelera, como si los amiloides inoculados participaran tambien de la patogenesis (68). Ahora bien: ¿no se tratara de autenticos priones defectivos, en forma análoga a lo que ocurre con retrotransposones que son retrovirus defectivos y con muchos virus?  es decir, proteínas que en el pasado fueron priones pero que perdieron la capacidad autorreplicativa y de transmisión, conservando su capacidad de formar amiloides y causar enfermedad. Si esto fuera cierto, tal vez con algun prion helper sintetizado artificialmente se podría inducir el comportamiento prionico de estas supuestas proteínas defectivas (esto que digo es análogo a tener un virus defectivo en una celula que no puede replicarse e infectar con un virus “helper que aporta la maquinaria de replicación que le falta al otro virus, con lo cual se va  a poder replicar).  De hecho, en algunos casos es posible acelerar una enfermedad.

Resulta evidente que los mecanismos por los cuales los priones ejercen su accion deleterea sobre las células de mamíferos son complejos y permanecen en gran parte en el misterio.  Basicamente todos involucran la apoptosis pero mediante diversas vias parcialmente conocidas. Otras más seguramente quedan por descubrir.  Se puede destacar tambien la semejanza entre la patogenia de los priones y la de ciertos virus que afectan el sistema nervioso en ratones. En ratón existen retrovirus que inducen la respuesta UPR a traves de la acumulación y agregación de proteínas virales de la capside en el retículo a las cuales se une Bip (110).  El mecanismo de toxicidad de prpsc en el retículo no esta determinado totalmente sino que la evidencia experimental apunta a que es el mismo que ocurre en el retrovirus murino. Resulta intrigante que dicho virus produce una enfermedad que se asemeja muchísimo a las TSEs, incluyendo vacuolización de las neuronas. Esto podria ocurrir mediante el mismo mecanismo  que la espongiosis de origen prionico, involucrando acuoporinas.

La semejanza entre el virus y el prion podria ser simplemente producto de la casualidad o tal vez el resultado de la evolución convergente, pero otra posibilidad apunta aun origen viral del gen PRP. Dicho gen podria provenir de un retrovirus ancestral que infectaba a algun vertebrado del cual derivaron todos los vertebrados, que actualmente tienen el gen. El virus se habría insertado en el genoma del vertebrado en la línea germinal mediante su mecanismo de transposición y habría quedado incorporado. Actualmente se piensa que mucho del DNA presente en los organismos deriva de retrovirus. Existen retrovirus endógenos, esto es, retrovirus que perdieron genes que requerian para multiplicarse, por lo que quedan latentes en el genoma y van acumulando mutaciones, ademas puede ocurrir duplicación y divergencia de sus secuencias. Luego, cada duplicacion evoluciona por separado. Originalmente, el gen PRP pudo pertenecer al genoma de un retrovirus y codificar una prpc perteneciente a la capside viral que por alguna razon fue conservada y actualmente es producida por las celulas de vertebrados actuales. La evidencia a favor de algunas de las funciones propuestas para prpc es fuerte, pero aun asi, hasta ahora nadie demostro la función in vivo sin que haya ninguna duda, a lo mejor, la dificultad para encontrar dicha función y determinar que realmente es esa la funcion in vivo se debe a que prpc cumpliría una funcion redundante, suplida por otras proteínas. Pero tal vez la imposibilidad de determinar la función se deba a que no la tiene más alla de servir para la propagacion de prpsc, pero Prpc bien podria derivar de una proteina viral y seguir siendo sintetizada en al celula si tenerla proporcionara alguna ventaja adaptativa, por ejemplo interferir con la multiplicación de virus y esa podría ser su función actual. Dentro del ciclo de multiplicación de un virus puede estar involucrado en muchos casos la síntesis de proteínas de la cápside que se translocan al retículo y siguen la via secretoria al igual que prpc. Si prpc derivara de proteínas virales ancestrales podria haber mantenido cierta semejanza con proteínas virales actuales y de esta manera interferir con la replicación viral a traves de la interaccion con las proteínas virales en el lugar en que se ensambla la capside (en algun punto de la ia secretoria, puede ser el retículo, el Golgi o las vesículas endocíticas). La interacción con proteínas de la cápside que se estan ensamblando podria resultar en la producción de cápsides defectivas que no incorporan el genoma viral, inhibiendo asi la infección, se trataría de un sistema de defensa. De todos modos esto es hipotético y no concuerda bien con la evidencia a favor de funciones relacionadas con transduccion de señal, estrés oxidativo y metabolismo del cobre. Harian falta nuevos estudios para aclarar este punto.

Los priones podrían en principio ser importantes en la adaptación de sus hospedadores al medio ya que tienen la capacidad de propagar una conformación proteica en toda la población de dicha proteína dentro de los individuos de una misma especie en forma altamente eficiente y relativamente rápida una vez que se encuentran dentro de la célula blanco, y también debido a que cada conformación tiene una información intrínseca distinta. Pueden formar parte de mecanismos de regulación y adaptación en los que inicialmente todas las proteínas de un mismo tipo tienen la misma conformación dentro de una población dada de cierta especie hospedadora y estan cumpliendo una funcion determinada. Ante algún cambio en el ambiente que resulta perjudicial para la especie, en algunos individuos podria cambiar espontáneamente la conformación del prion, que entonces se va a propagar entre los individuos de la población, en algunos casos, la nueva conformación podria favorecer la adaptación al cambio, en cambio, otras nuevas conformaciones podrían ser perjudiciales (por ejemplo el caso de prpsc). Cualquier conformación del prion que surja y se comporte como las formas infecciosas descriptas en este trabajo, se va a propagar y va a reeemplazar a todas las moléculas de otra conformación que encuentre, por esto, en principio tanto el prion que confiere ventaja adaptativa y el que resulta perjudicial se van a propagar, pero el perjudicial va a estar sujeto a selección natural negativa, es decir va a tender a perderse pues los organismos que lo poseen seran menos viables y dejaran menos descendencia. Esto no se cumple en el caso de prp porque es un prion lento, las enfermedades que produce tienen un largo periodo de incubación de decadas y no impiden que el individuo portador deje descendencia. La enfermedad suele desencadenarse después del fin de la edad reproductiva.

En cambio un prion beneficioso, va ser mantenido en la población por selección natural y se va propagar mas fácilmente pues sus portadores tendran un fitness mayor.

Otra funcion de los priones se relaciona al mantenimiento de la memoria. En un molusco del genero Aplysia, se encontró una proteína con características típicas de prion denominada elemento de poliadenilación citoplasmática proteico (CPEB). Esta proteína activa mRNAs latentes (mensajeros que fueron sintetizados pero que no se estan traduciendo) mediante la elongación de sus colas de poliA y esto esta asociado a estabilidad sinaptica de largo alcance (conservación de la memoria). CPEB existe en dos conformaciones y la conformación que se encarga de elongar los mensajeros y de esta forma participar del proceso de mantenimiento de la memoria es la conformación infecciosa que se propaga como todo prion (convirtiendo moléculas de la otra conformación, que es inactiva en mas moléculas de si misma) (116). Seguramente existen muchísimos casos mas como este. Los priones son proteínas ubicuas de los que actualmente se sabe muy poco y puede que sean mas importantes de lo que se piensa. Tal vez se trata de otro caso similar al del RNA, del que hace algunos pocos años nadie imaginaba la cantidad de funciones regulatorias que tiene a traves del mecanismo de RNAi.

## Agradecimientos

Este trabajo fue realizado gracias a una iniciativa de los docentes de quimica biologica II, materia dictada en la Facultad de Ciencias Exactas y Naturales de la Universidad de Buenos Aires.